\documentclass[graybox]{svmult}

\usepackage{type1cm}        
\usepackage{makeidx}         
\usepackage{graphicx}        
\usepackage{multicol}        
\usepackage[bottom]{footmisc}

\usepackage{dsfont}
\usepackage{bigints}
\usepackage{tikz}
\usepackage{tikz-3dplot}
\usepackage[breaklinks=true,colorlinks,citecolor=blue,linkcolor=blue,urlcolor=blue]{hyperref}

\newcommand{\beq}{\begin{eqnarray}}
\newcommand{\eeq}{\end{eqnarray}}

\newcommand{\real}{{\sf I}\kern-.12em{\sf R}}
\newcommand{\comp}{{\sf I}\kern-.50em{\sf C}}
\newcommand{\unity}{{\sf I}\kern-.54em{\sf 1}}

\DeclareMathOperator{\Vol}{Vol}

\newcommand{\mcTau}{\mathcal{T}}
\newcommand{\YMconfs}{\mathcal{C}_{G}\big( \mcTau \big)}

\usepackage{newtxtext}       %
\usepackage{newtxmath}       

\makeindex             

\begin{document}

\title*{Spectral Observables and Gauge Field Couplings in Causal Dynamical Triangulations}
\author{Giuseppe Clemente and Massimo D'Elia}
\institute{Giuseppe Clemente \at Deutsches Elektronen-Synchrotron DESY,\\ Platanenallee 6, 15738 Zeuthen, \email{giuseppe.clemente@desy.de}
\and Massimo D'Elia \at Dipartimento di Fisica dell'Universit\`a di Pisa and INFN
- Sezione di Pisa,\\ Largo Pontecorvo 3, I-56127 Pisa, Italy. \email{massimo.delia@unipi.it}} %
\maketitle

\abstract{In the first part of this Chapter, 
we discuss the role of spectral observables, 
describing possible ways to build them 
from discretizations of the Laplace--Beltrami 
operator on triangulations, 
and how to extract useful geometric information. 
In the second part, 
we discuss how to simulate the composite 
system of gauge fields coupled to CDT for generic groups and dimensions, showing
results in some specific case and 
pointing out current challenges.}

\section{Introduction}
\label{sec:intro}

One of the most promising results
of pure-gravity CDT in 4D is that
it appears to be non-perturbatively 
renormalizable in a Wilsonian renormalization group sense, 
i.e.~there exist second-order critical points 
which are candidates for extracting
continuous physics (see Chapters 1 and 10 of this Section~\cite{Ambjorn_chapt1,Gizbert-Studnicki_chapt10}). 
However, there is still an urge to 
identify a possibly complete set of physically meaningful 
observables to characterize all relevant features of 
the geometries under investigation.

The approach we followed in this respect is based on
\emph{spectral methods}, which are a set of techniques
involving the analysis of eigenvalues and eigenvectors 
of discretizations of the Laplace--Beltrami (LB) operators 
associated with spaces of functions on a manifold.
One of the advantages of the spectral decomposition of 
the LB operators into eigenvalues and eigenvectors is 
contained in their hierarchical nature,
which allows us to consistently separate large-scale features
from short-scale ones; 
in general, the spectrum of a LB operator
on a manifold M identifies a set of characteristic length 
scales, as we argue in Section~\ref{sec:spectral_methods}, while features 
like directionality, localization, and all the remaining 
metrical information can be extracted from eigenvectors,
which take the role of waveforms (or diffusion) modes.

Another direction that we investigated in order to 
extract meaningful physical properties 
of the system from CDT simulations is the problem
of minimally coupling gravity and 
Yang-Mills gauge fields, possibly 
including also fermionic matter. 
Not only this is theoretically 
important in the search for a complete 
and self-consistently renormalizable 
quantum theory of gravity in the continuum, 
since the composite system could possess different
critical properties than the pure-gravity one,
but also because this allows an easier connection 
between the theory and possible phenomenological
results from quantum cosmological observations, 
provided these would be available at some point.
In Section~\ref{sec:gf}, we describe 
the algorithmic strategies that we use to implement 
this minimal coupling, and the challenges that 
appear for higher dimensions and gauge groups,
both algorithmic and regarding the definition 
of meaningful observables. 

\section{Spectral methods} 
\label{sec:spectral_methods}
The relationship between the geometric properties of a manifold and the spectral decomposition
of its associated Laplace--Beltrami (LB) operators is well known in literature~\cite{reuter_cad,reuter_dna,lapl_embedding},
and spectral analysis, i.e.~the study of eigenvalues and eigenvectors of relevant operators of the model under study,
shows a wide spectrum of applications through science and beyond.
In this Section, we first introduce some basics regarding the physical role and interpretation
of spectra and eigenvectors, followed by a graph discretization of the LB operator with some toy examples. 
Then we show some numerical results for CDT and discuss other possible discretizations.

\subsection{General properties of spectra and eigenvectors}
\label{subsec:spectraGeneral}

In order to understand the physical content of the LB spectral decomposition, it is useful to start
from the example of diffusion processes, for simplicity on boundaryless manifolds $\mathcal{M}$ (either smooth or piecewise flat, as the ones appearing in dynamical triangulations). On a chart with coordinates $\mathbf{x}$,
and with diffusion time $t$ (which is not a physical time), the heat equation takes the form 
\begin{equation}\label{eq:heat}
\partial_t u(\mathbf{x},t)= D^\mu D_\mu u(\mathbf{x},t),
\end{equation}
where $D_\mu$ and $-D^\mu D_\mu$ are respectively the covariant derivative and the LB operator on $\mathcal{M}$.
One then can proceed by expanding solutions on the basis of an orthonormal set of LB eigenfunctions $\phi_k(\mathbf{x})$ for each $t$
\begin{equation}
u(\mathbf{x},t)=\sum_{k=0}^{\infty} c_k(t) \phi_k(\mathbf{x}),
\end{equation}
and substitute into Equation~\eqref{eq:heat} to obtain 
\begin{equation}
\frac{d}{dt} c_k(t)=-\lambda_k c_k(t) \quad \longrightarrow \quad c_k(t)=c_k(0) \; e^{-\lambda_k t},
\end{equation}
with $\lambda_k$ the eigenvalues of the LB operator associated to $\phi_k(\mathbf{x})$.
The coefficients $c_n(0)$ are fixed by the initial condition $u(\mathbf{x},0)$. 
It is straightforward to show that the $L_2$ norm (weighted by the metric term $\sqrt{g(\mathbf{x})}$) of
$u(\mathbf{x},t)$ is constant in diffusion time $t$, so we can interpret a solution with norm 1 as a probability distribution.
In the case when one starts the diffusion process from a distribution where all 
probability is concentrated at one point $\mathbf{y}$, the fundamental solution to the heat equations 
\begin{align}\label{eq:heatkerneq}
G(\mathbf{x},\mathbf{y};0)&=\frac{1}{\sqrt{\det(g(\mathbf{x}))}} \delta^d(\mathbf{x} - \mathbf{y}),\\
\partial_t G(\mathbf{x},\mathbf{y};t) &=D_x^2 G(\mathbf{x},\mathbf{y};t),
\end{align}
is the so-called \emph{heat-kernel}, and it can be written either in terms of 
spectrum and eigenvectors of the LB operator, or as an early (diffusion) time expansion, for any $\mathbf{x}$ and $\mathbf{y}$, reads as follows~\cite{heatk}:
\begin{equation}\label{eq:hsolseries}
    G(\mathbf{x},\mathbf{y};t) =\sum_{k=0}^{\infty} e^{-\lambda_k t} \phi_k(\mathbf{x})\phi_k(\mathbf{y}) \sim \Big[ \frac{e^{-d_g^2(\mathbf{x},\mathbf{y})/4t}}{{(4 \pi t)}^{d/2}} \Big] \sum_{n=0}^{\infty} a_n(\mathbf{x},\mathbf{y}) \; t^n 
\end{equation}
where we denote by $d_g(\mathbf{x},\mathbf{y})$ the \emph{geodesic distance} 
between  the points $\mathbf{x}$ and $\mathbf{y}$. 
For a $d$-dimensional flat space, the term is square brackets of 
Eq.~\eqref{eq:hsolseries} is the only one present, because 
$a_0(\mathbf{x},\mathbf{y})$ is the only non-zero term in the series. 
Furthermore, for non-flat but smooth manifolds, the small $t$ regime probes 
length scales of the order $d_g(\mathbf{x},\mathbf{y}) \sim \sqrt{t}$, 
due to the exponential falloff.
This is connected to a property of the spectral version of the heat-kernel, 
appearing in the left expressions of Eq.~\eqref{eq:hsolseries},
which consists in how eigenmodes $\phi_k(\vec{x})$ 
with relatively larger eigenvalues $\lambda_k$ 
are exponentially suppressed with respect to smaller eigenvalues. 
Indeed, this shows a key feature of the spectrum 
(even more evident if one considers
the wave equation): 
the lowest part of the spectrum is associated to the slowest diffusion modes, 
while the highest part to the fastest. Since the eigenmodes of the LB operator
are also solutions to the wave equation, a direct correspondence can be made between
eigenvalues and the (inverse squares of the) wavelengths: in these terms, 
the lowest part of the spectrum is associated to the longest wavelengths\footnote{There is a 
disclaimer on the wave interpretation of
eigenmodes: since the typical geometries we consider are random, 
the eigenmodes often exhibit an Anderson-like localization 
behavior~\cite{Anderson:1958vr,qcd_anderson_multifractal}, instead of being long-range.} 
while the highest eigenvalues are associated to the smallest ones,
which, being related to the ultraviolet scales,
can be neglected in the investigation of continuum physics.
Another relevant quantity, related to the heat-kernel $G(\mathbf{x},\mathbf{y};t)$,
is the average \emph{return probability}~\cite{fractals_havlinbook,edt_spectral_dim,cdt_spectral_dim,diffproc}, defined as:
\begin{equation}\label{eq:retpr}
P(t):=\frac{1}{\Vol(\mathcal{M})} \int d^d \mathbf{x} \sqrt{\det(g(\mathbf{x}))} \; G(\mathbf{x},\mathbf{x};t),
\end{equation}
which can be analogously expanded in $t$ as:
\begin{equation}\label{eq:retprseries}
P(t) =\frac{1}{\Vol(\mathcal{M})} \sum_{k=0}^{\infty} e^{-\lambda_k t} \sim \frac{1}{{(4 \pi t)}^{d/2} \Vol(\mathcal{M})}\sum_{n=0}^{\infty} A_n \; t^n.
\end{equation}
The coefficients $A_n$ are related to a useful hierarchy of geometric 
quantities such as volume ($A_0 = \Vol(\mathcal{M})$), total curvature ($A_1 = \frac{1}{3} \int_{\mathcal{M}} R$), and other diffeomorphism invariant 
scalars built from contractions of higher powers of the Riemann tensor~\cite{heatk,heatrace_coeffs}.
On a flat space $\mathbb{R}^d$, the return probability reduces to a power law behavior $P(t)={(4\pi t)}^{-\frac{d}{2}}$ (with all coefficients 
$A_{i\geq 1}=0$),  as expected from scale-invariance. For a (non-pathologically) curved but smooth manifold of dimension $d$, 
the leading diffusion behavior at small 
diffusion times acts is the same as the one of flat space with the same dimension. 
Therefore, in the case of a smooth manifold $\mathcal{M}$, 
just by using the LB spectrum, 
or equivalently by the return probabilities of diffusion processes 
at different diffusion times, 
it is possible to extract information about the dimension of the space
\begin{equation}\label{eq:diffspdim}
d=-2 \lim_{t\to 0^+} \frac{d \log P(t)}{d \log t}.
\end{equation}
The geometries appearing in (C)DT are far from smooth,
and that means that a small diffusion time extrapolation as the 
one shown in Eq.~\eqref{eq:diffspdim} is not really meaningful.
However, one is not interested in the ultraviolet behavior, 
which is the one affected by discretization artifacts,
but is sufficient to characterize geometries locally in a mesoscopic sense,
i.e.~for intermediate scales. Indeed, by universality arguments, 
the continuum limit behavior should be affected only by relevant operators which 
are insensitive to the finer details of the regularization.
As we show in the next section, it is possible to extend the regime of validity
of Eq.~\eqref{eq:diffspdim} up to finite diffusion times (instead of evaluating in 
the limit $t\to 0^+$) by defining the so-called \emph{spectral dimension}, 
which can be evaluated at different diffusion times (or equivalently, length scales).

\subsubsection{Weyl's law, spectral dimension and connectivity}\label{subsubsec:weylslaw}

Let us consider again a $d$-dimensional smooth manifold $\mathcal{M}$ 
without boundaries and with spectrum $\sigma(\mathcal{M})$ of the LB operator.
The count of eigenvalues below a spectral radius $\lambda$
can be written as 
\begin{align}\label{eq:countingdef}
n(\lambda)\equiv \sum_{\lambda_k\in \sigma(\mathcal{M})} \theta(\lambda-\lambda_k) = \int_{0}^{\lambda} \rho(\lambda^\prime) d\lambda^\prime,
\end{align}
where we also defined the spectral density $\rho(\lambda)\equiv \frac{dn}{d\lambda}=\sum_k \delta(\lambda-\lambda_k)$. 
As clear from its definition in terms of eigenvalues, 
is possible to write the return probability, 
shown in Eq.~\eqref{eq:retprseries}, 
as the Laplace transform of the spectral density 
$P(t) = \mathcal{L}[\rho(\lambda)](t)$. 
Likewise, the spectral density can then be written as 
an inverse Laplace transform of the return probability\footnote{
This comes from the observation that 
the inverse Laplace transform of $\frac{1}{t^\alpha}$ is $\mathcal{L}^{-1}[t^{-\alpha}](\lambda)=\frac{\lambda^{\alpha-1}}{\Gamma(\alpha)} \theta(\lambda)$ 
and that $n(\lambda)$ is the integral of $\rho(\lambda)$.} 
$\rho(\lambda) = \mathcal{L}^{-1}[P(t)](\lambda)$ which, 
due to the leading small diffusion time behavior $P(t)\propto t^{-d/2}$, 
establishes a connection with another important result of spectral geometry called 
\emph{Weyl's law}~\cite{weylslaw_1,weylslaw},
describing the asymptotic behavior of the eigenvalue counts 
$n(\lambda)$ at large spectral radius $\lambda$:
 \begin{equation}\label{eq:weyl_law}
 n(\lambda) \sim \frac{\omega_d V}{(2 \pi)^d} \lambda^{d/2}, 
 \end{equation}
where $\omega_d$ is the volume of the $d$-dimensional ball of unit radius
and $V\equiv \Vol(\mathcal{M})$ indicates the volume of the manifold.
The leading dependence in expression in Eq.~\eqref{eq:weyl_law} 
can be used to build an alternative definition of the spectral dimension,
which we call \emph{effective dimension}~\cite{LBseminal}, as a quantity that is in general 
running at different energy scales $\lambda$:
\begin{equation}\label{eq:elevspdim}
d_{EFF}(\lambda) \equiv 2\frac{d \log (n/V)}{d \log \lambda}.
\end{equation}
As mentioned in the previous Section, another definition of dimension, called \emph{spectral dimension}~\cite{edt_spectral_dim,cdt_spectral_dim,diffproc},
can instead be defined in terms of return probability and diffusion times
as an extension of the regime of validity of Eq.~\eqref{eq:diffspdim} 
to finite diffusion times $t$, namely:
\begin{equation}\label{eq:spdim_extended}
D_S(t) \equiv -2\frac{d \log P}{d \log t}.
\end{equation}
It should be evident at this point  that the two definitions 
in Eq.~\eqref{eq:elevspdim} and~\eqref{eq:spdim_extended} 
are intimately connected through Laplace transform, which treats
the diffusion time $t$ and eigenvalue ``energy'' $\lambda$ 
as variables dual to each other.

Before showing the results of the application in the CDT case, 
it is useful to see these definitions in action and get some intuition
on their general behavior using some simplified toy examples 
on discretized geometries, for which the smoothness condition is not available,
but where one can nevertheless extract geometric information 
at the mesoscopic scales.

\begin{figure}[!t]
\centering
 	\includegraphics[width=0.49\linewidth]{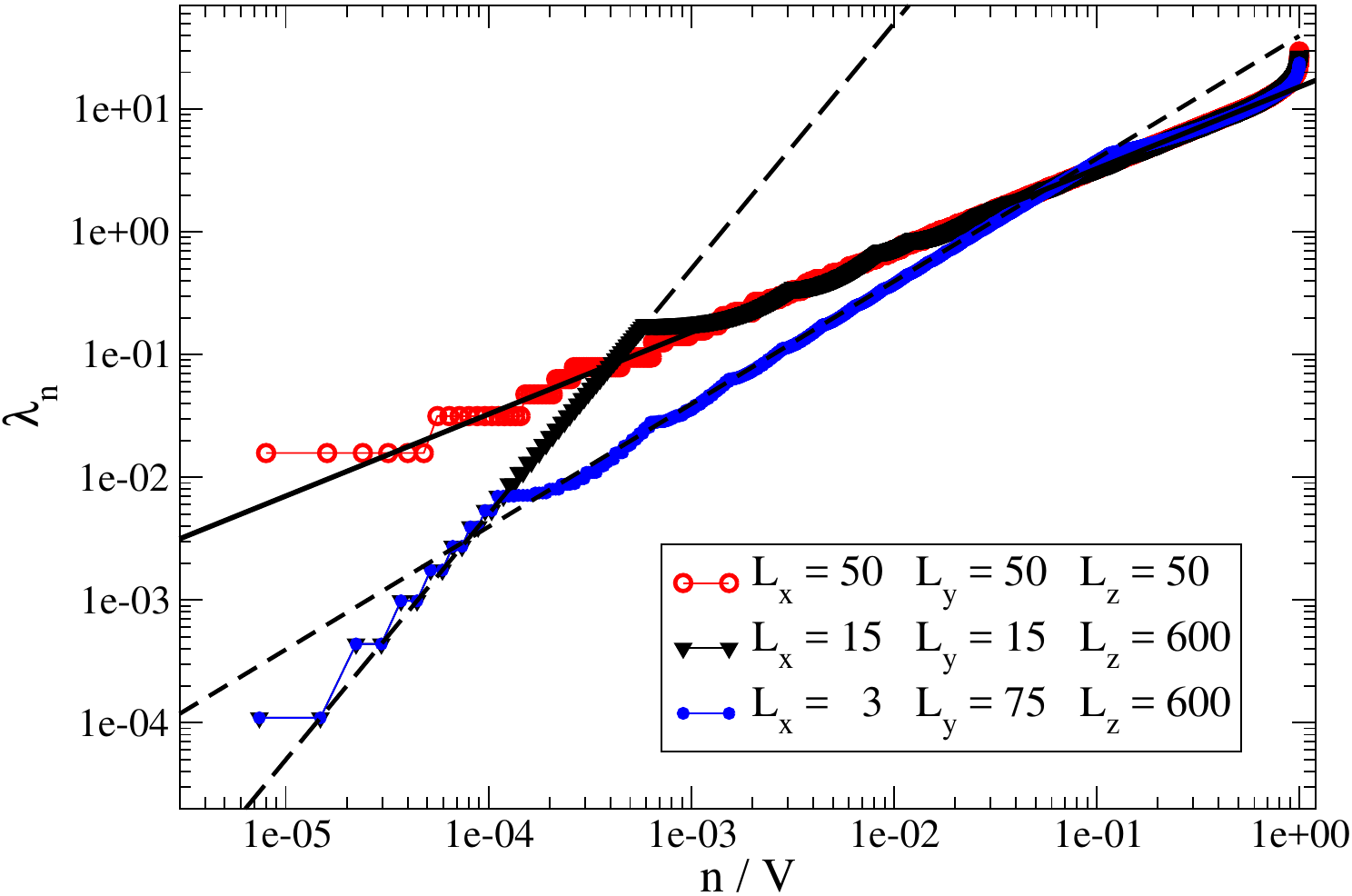}
\includegraphics[width=0.49\textwidth]{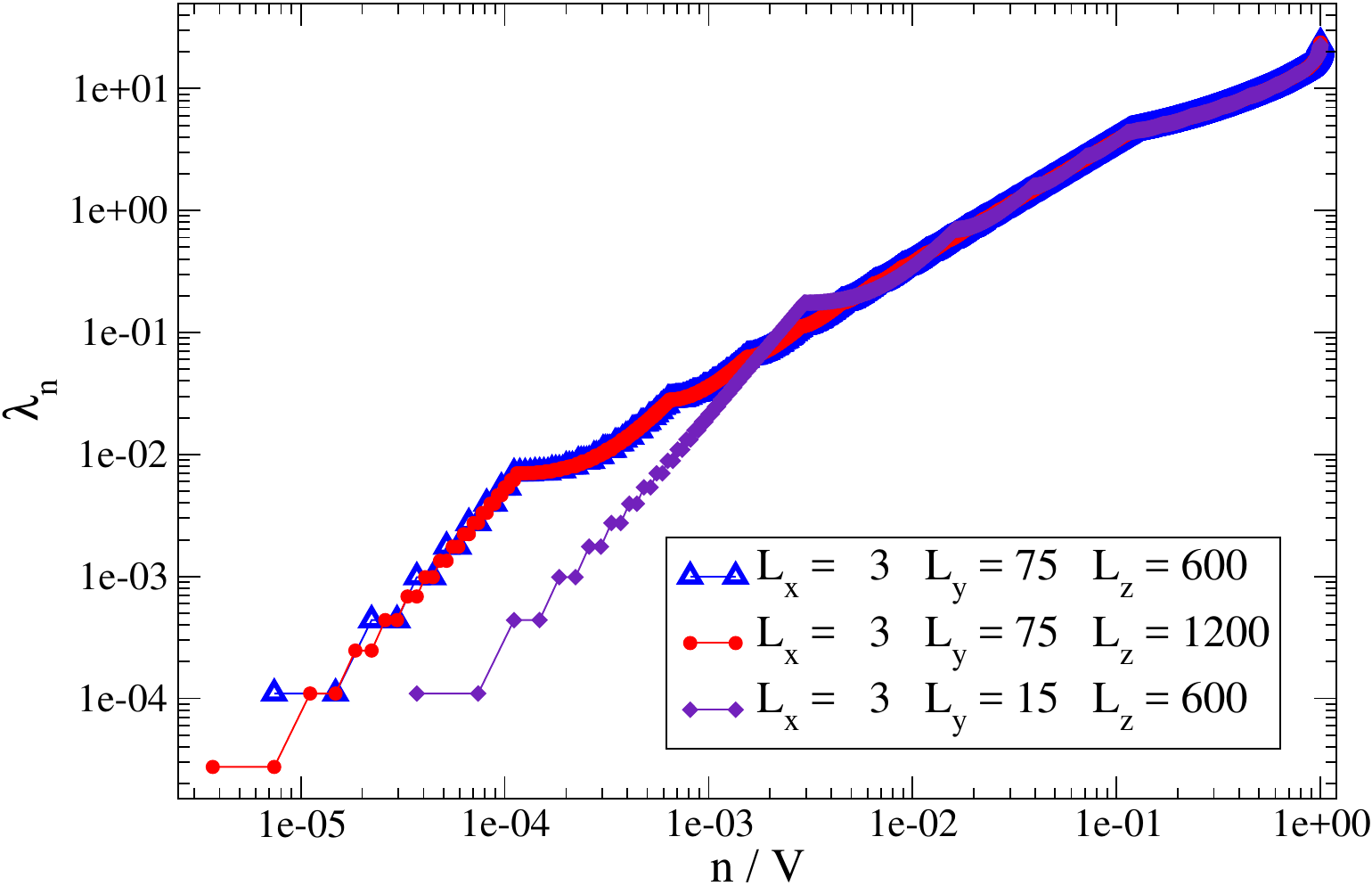}
 	\caption{Plots of $\lambda_n$ against its volume-normalized order $n/V$, for a hypercubic lattice with periodic boundary conditions (i.e.~toroidal) and different combinations of sizes $L_i$ for each direction. 
 	The straight continuous line follows the exact Weyl scaling (see Eq.~\eqref{eq:weyl_law}) 
  expected from a $d=3$ dimensional space, while the dashed straight lines correspond 
  to effective Weyl scalings for effective dimensions $d=2$ and 1, 
  which dominate the largest scale behavior.}
\label{fig:toymodel_1}
\end{figure}

Let us consider a finite difference discretization of a 3-dimensional torus
as a regular cubic lattice, with respectively 
$L_x$, $L_y$ and $L_z$ sites along three orthogonal directions.
Eigenmodes are periodic plane waves $\phi_k(n_x,n_y,n_z) \propto e^{i (k_x n_x +k_y n_y +k_z n_z)}$
with wave number $\vec k = (k_x, k_y, k_z)$,
where $k_i = 2 \pi\, m_i / L_i$ and $m_i$ integers such that $-L_i/2 < m_i \leq L_i/2$.
Using $\vec{m}$, the spectrum can be written as
\begin{equation}
\lambda_{\vec m} = \sum_{\mu=x,y,z} 4 \sin^2\Big(\frac{\pi m_\mu}{L_\mu}\Big),
\end{equation}
but, for our applications, it is more useful to relabel them with a single integer label $n$ and 
in non-decreasing order $\lambda_{n+1}\geq \lambda_{n}$. This allows us to use 
Eq.~\eqref{eq:elevspdim} to extract the effective dimension at different scales from the
slope in log-log plots of $\lambda_n$ versus the \emph{volume-normalized order} $\frac{n}{V}$ 
(zero-mode excluded).
Figure~\ref{fig:toymodel_1} 
shows, for different combinations of
sizes $L_i$, how there is a quite well-defined scale separation between different eigenvalue ranges
with different slopes and therefore different effective dimensions.
In particular, notice in the right panel of Figure~\ref{fig:toymodel_1} how the triangle and circle dots, 
associated with tori which differ only by $L_z$ (and therefore the total volume),
exhibit a collapse of their trends and the same slope behavior where transitions between different scaling dimensions happen
at the same point in $(\lambda,\frac{n}{V})$, 
which determines also the smallest eigenvalue, or equivalently, the largest linear size that can be reached. 
This shows that, using the volume-normalized order 
$\frac{n}{V}$ instead of just the eigenvalue order label $n$ is essential to compare 
the spectra from triangulations with different volumes, 
as we do in Section~\ref{subsec:spectraCDT}.

This example highlights also another important point: 
since the smallest non-zero eigenvalue, called \emph{spectral gap},
sets the largest length scale of manifolds, 
it should vanish in the thermodynamical limit, i.e.~when the volume diverges.
In particular, we expect it to follow a power-law behavior set by 
\begin{align}\label{eq:gapvolume_relation}
\lambda_1 \propto V^{-\frac{2}{d_{EFF,\text{low}}}},
\end{align}
where $d_{EFF,\text{low}}$ is the effective dimension in the lowest region of the spectrum 
(large scales). However, in Section~\ref{subsubsec:bifurc} we show that
triangulations in the $B$ phase, and some slices in the bifurcation phase $C_b$ possess 
a gap that does not vanish in the thermodynamical limit. This is related to the observation
that geometries in those cases are typically highly connected, and the effective dimension 
at larger scales is diverging, which is consistent with the relation in Eq.~\eqref{eq:gapvolume_relation}.
The intimate relation between spectral gap and connectivity can be made more explicit. 
Indeed, a measure of connectivity for a compact Riemannian manifold $\mathcal{M}$ is 
encoded in the \emph{Cheeger isoperimetric constant} $h(\mathcal{M})$, 
defined as the minimal area of a hypersurface $\partial A$ which bipartitions $\mathcal{M}$ 
into two disjoint pieces $A$ and $\mathcal{M}\setminus{A}$ in the most balanced way 
\begin{equation}
h(\mathcal{M}) \equiv \inf \frac{\Vol(\partial A)}{\Vol(A) \Vol(\mathcal{M}\setminus{A})} \, ,
\end{equation}
where the infimum is taken over all possible connected submanifolds $A$.
The connectivity, in the form of the Cheeger constant, is bounded by the spectral gap through the \emph{Cheeger inequality}~\cite{cheeger}
\begin{equation}\label{eq:Cheeger_ineq}
h(\mathcal{M}) \leq 2 \sqrt{\lambda_1}\,. 
\end{equation}

Up to this point, we discussed general properties of spectra without
referring to specific discretizations of the LB operator.
In the next Section we introduce a graph discretization
and show some results of the application of spectral analysis
in CDT.

\subsection{Graph discretization of the Laplace--Beltrami operator}\label{subsec:lap_dg}
\begin{figure}[ht]
	\centering
	\includegraphics[width=0.9\linewidth,trim={3em 3em 3em 3em},clip]{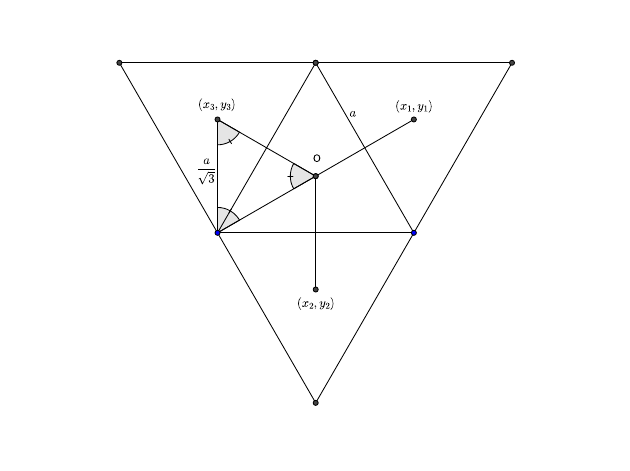}
	\caption{Sketch of the dual graph discretization of the LB operator in 2D.}
    \label{fig:dgcalc}
\end{figure}

Here we discuss a specific class of discretizations 
of the Laplace--Beltrami operator on simplicial manifolds, namely, graph discretizations.
In particular, we focus on the discretization as Laplace matrix on the graph dual to the triangulation,
where elementary blocks of spacetime volumes, i.e. simplexes, are represented as nodes, 
while the adjacency relations between them are represented by links.
In order to show how this discretization directly connects with
functions on the continuous piecewise flat manifold,
let us consider a scalar function $f(x)$ on a $d$-dimensional chart
formed by a $d$-simplex $\sigma_0$ and all its adjacent simplexes $\{\sigma_k\}_{k=1,\dots,d+1}$ , assuming equilateral lengths 
of size $a$ for each simplex (Figure~\ref{fig:dgcalc} illustrates the $2$-dimensional case).
Since the space inside the chart is flat everywhere, one can Taylor expand
around the barycenter $\vec{x}_0$ of $\sigma_0$ and with respect to the
barycenters $\{\vec{x}_k=\vec{x}_0 + \delta \vec{x}_k\}_{k=1,\dots,d+1}$ of the adjacent simplexes as follows:
\begin{align}
    f(\vec{x}_0+\delta \vec{x}_k) = f(\vec{x}_0) + \delta x^i_k \partial_i f(\vec{x}_0) + \frac{1}{2} \delta x^i_k \partial_i\partial_j f(\vec{x}_0)\delta x^j_k + O({|\delta \vec{x}_k|}^3).
\end{align}
In the equilateral case, the magnitude of the displacements $\delta \vec{x}_k$ are all proportional to the side lengths $a$,
i.e., $\delta \vec{x}_k = \frac{a}{\sqrt{d+1}} \hat{v}_k$ for some
unit vectors $\hat{v}_k$. 
Furthermore, the unit vectors all sum up to the zero vector $\sum_{k=1}^{d+1} \hat{v}_k = \vec{0}$,
so one can eliminate the gradient term by summing over all displacements $\delta \vec{x}_k$.
The best local approximation of the LB operator, using the function evaluation 
at the barycenters is then given by\footnote{Summing over all displacements in the equilateral case 
results in $\sum_{k=1}^{d+1} \delta x^i_k \delta x^j_k=\frac{a^2}{d+1}\sum_{k=1}^{d+1} v^i_k v^j_k= a^2 \delta_{i,j}$.}
\begin{align}
   -\Delta f(\vec{x}_0) = \frac{1}{a^2} \Big[(d+1)f(\vec{x}_0) - \sum_{k=1}^{d+1} f(\vec{x}_k)\Big] + O(a),
\end{align}
In this discretization, the space of scalar functions on the simplicial manifold can 
then be approximated by the set of values at the barycenter of each simplex 
$\{p_i\}_{i=0}^{|V|}$, i.e. $\vec{f}=(f(p_0),f(p_1),\dots,f(p_{|V|-1})) \in \mathbb{K}$, 
for some field $\mathbb{K}$, while the LB operator takes the form of a matrix,
called \emph{graph Laplacian}
\begin{equation}\label{eq:disclap}
-\triangle \rightarrow L\;= \; (d+1) \cdot \mathds{1} - A.
\end{equation}
where $A$ is the adjacency matrix, whose entry $A_{ij}$ is $1$ if the simplexes labeled 
with $i$ and $j$ are adjacent and $0$ otherwise. 

In the next Section, we show some results of the application of the 
dual graph Laplacian to investigate the geometric properties of CDT configurations.

\subsection{Spectral properties of CDT configurations}\label{subsec:spectraCDT}

\begin{figure}[ht]
  	\centering
  	\includegraphics[width=1\linewidth]{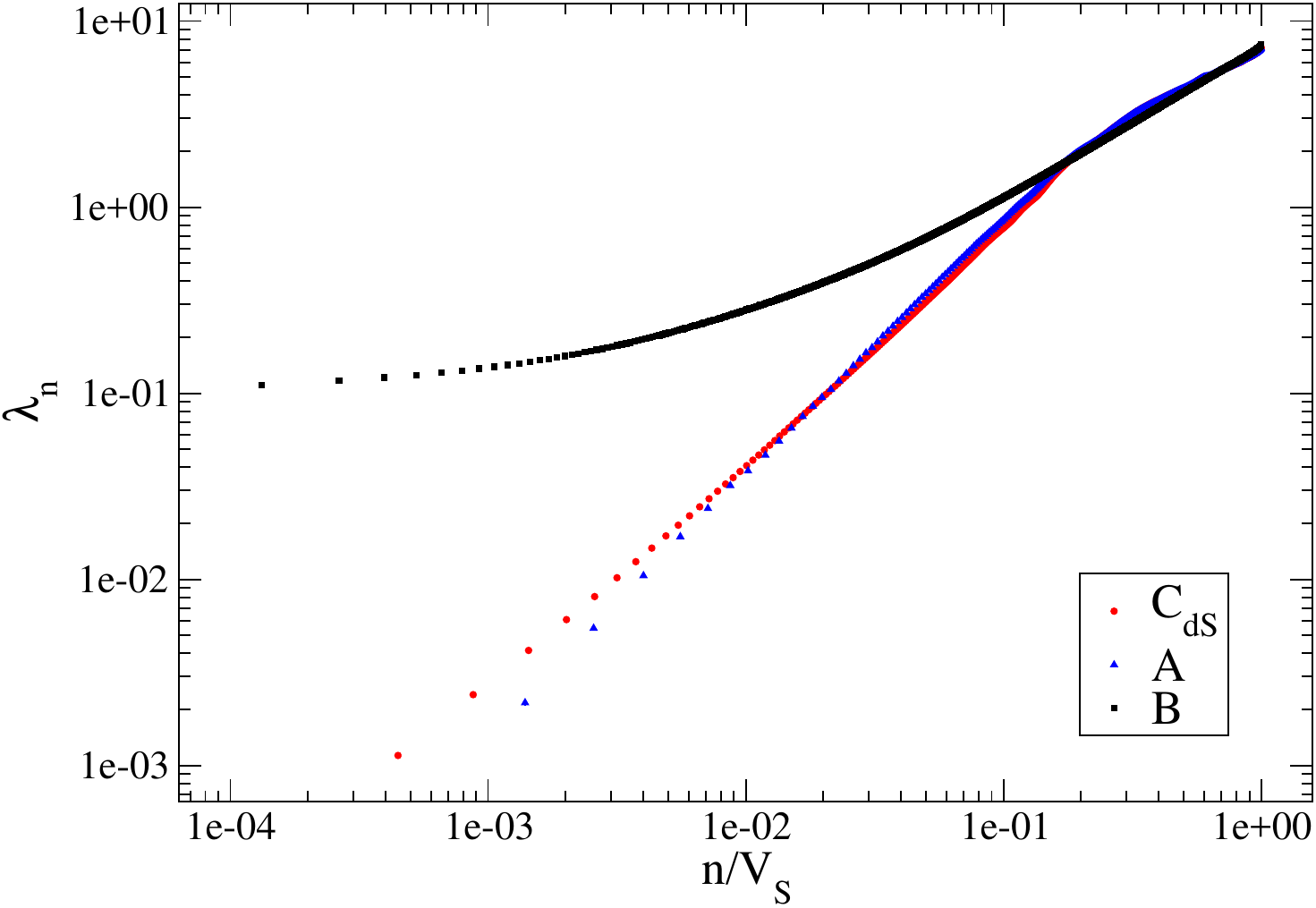}
  	\caption{
  		Averages of $\lambda_n$ versus $n/V_S$ computed in bins of $n/V_S$ with size $2/V_{S,max}$ for slices taken from configurations sampled deep into the $A$, $B$ and $C_{dS}$ phases. The volume is fixed to $V_{S,tot}=40k$ for configurations in $A$ and $C_{dS}$ phase, and to $V_{S,tot}=8k$ for configurations in $B$ phase. See Ref.~\cite{LBseminal} for details.}
  	\label{fig:averbin_CdS_A_B-lam-vs-k_V}
  \end{figure}

\begin{figure}[t]
    \includegraphics[width=0.9\columnwidth, clip]{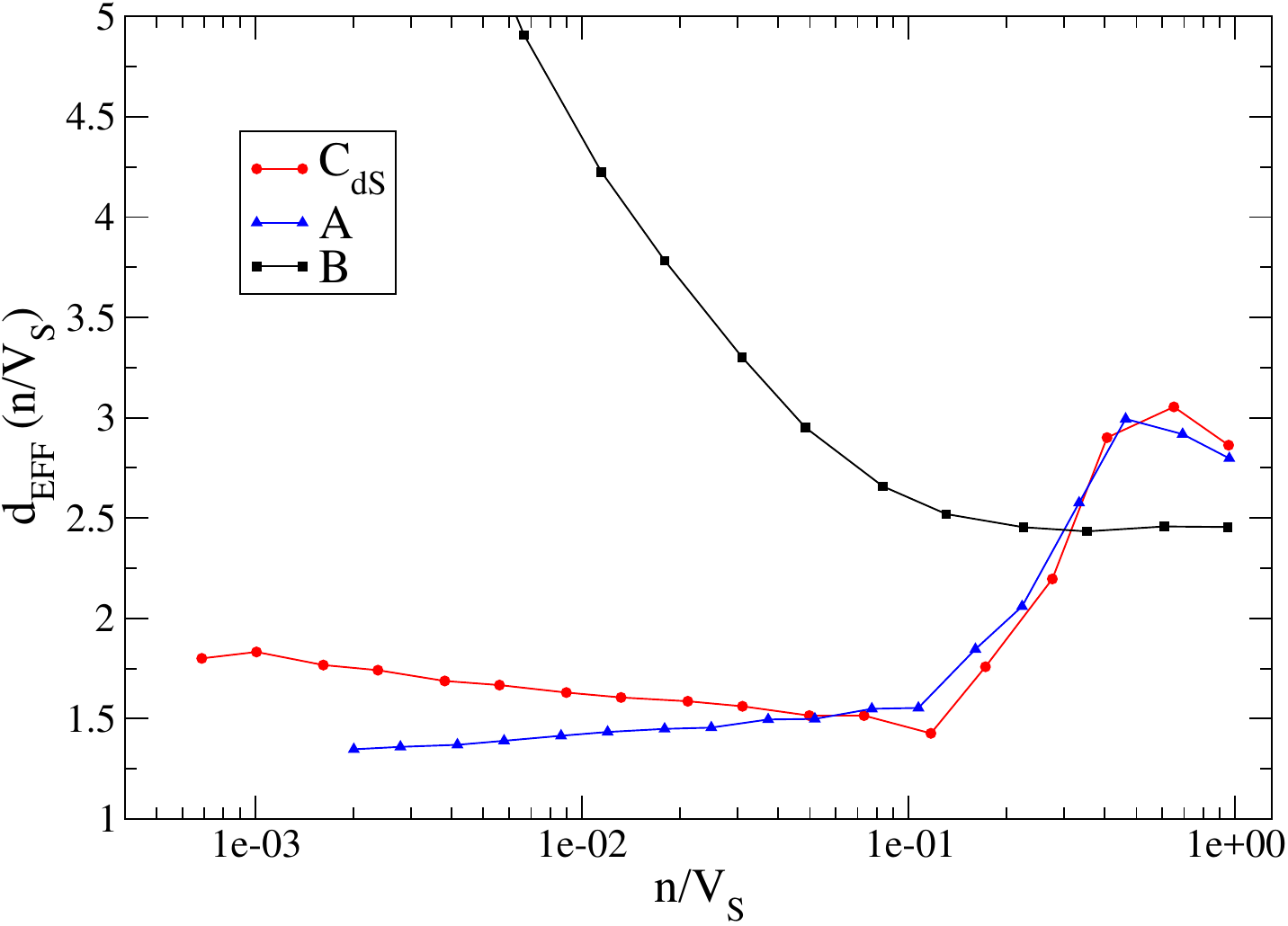}
    \caption{Running dimension obtained from the logarithmic 
slope of $\lambda$ vs $\frac{n}{V_S}$ curves (see Section~\ref{subsubsec:weylslaw} and Eq.~\eqref{eq:elevspdim}), computed over windows of different ranges of $n/V_S$ and for configurations sampled 
in phases $C_{dS}$, $A$ and $B$. The curve associated to the $B$ phase is diverging for $n/V_S \rightarrow 0$ (it is around 30 at 
$n/V_S \sim 10^{-4}$), but part of it has 
been omitted from the plot, to improve the readability of the curves obtained
for the other two phases. See Ref.~\cite{LBseminal} for details.}
    \label{fig:plot_CdS_A_B-Dk_V}
\end{figure}

In previous sections, we introduced the main tools for spectral
analysis. Here we show some qualitative and quantitative results
of their application.
As discussed in Chapters 1 and 10 of this Section~\cite{Ambjorn_chapt1,Gizbert-Studnicki_chapt10},
in 4-dimensional CDT there have been identified 4 phases, 
called $A$, $B$, $C_{dS}$, and $C_b$.
These can be characterized in the first place by their typical 
volume profiles, i.e.~the spatial volume per slice time $V_S(t)$.
The $A$ phase is considered unphysical since it exhibits small or 
absent correlation between adjacent slices, 
which is not observed in nature. $B$ phase is also considered 
unphysical, since all space volume is concentrated on a slice.
$C_{dS}$ and $C_b$ phases show instead spatial volumes 
more distributed in slice time and are therefore appealing 
for an investigation of the continuum limit, 
in particular at the transition between $C_{dS}$ and $C_b$ phases,
which appears to be second-order~\cite{cdt_secondord,cdt_secondfirst,new_phase_chars,cdt_newhightrans}. 
In the rest of this section and the next, 
we focus on the spectral properties of 
spatial slices of CDT configurations, since they capture the 
essential differences between phases, in particular for the 
$C_b$ phase, discussed separately in Section~\ref{subsubsec:bifurc}.
As done in Section~\ref{subsubsec:weylslaw},
we can extract the effective dimension~\eqref{eq:elevspdim}
from the slopes of $\lambda,n/V_S$ plots. 
Figure~\ref{fig:averbin_CdS_A_B-lam-vs-k_V} shows the 
binned average of the collapsed curves in $(\lambda,\frac{n}{V_S})$
for slices of some typical configurations, while
Figure~\ref{fig:plot_CdS_A_B-Dk_V} shows their corresponding 
effective dimension.
Notice how the single dominant slice in the $B$ phase configurations 
has a diverging effective dimension on long range. 
As argued in Section~\ref{subsubsec:weylslaw}, this is related to
a gapped spectrum with relatively large $\lambda_1$, 
associated to a large connectivity of the geometry: 
since the spectral gap sets the largest scale possible,
in $B$-phase slices, different regions are always quite close 
to each other, which accounts for the general geometry to be 
very connected. We are not interested in the properties of
$A$-phase spatial slices, which appear to be vaguely similar to
the one of $C$-phases slices, 
characterized by an almost constant 
effective dimension in a wide range of scales and with fractional 
value (the building blocks are 3-dimensional in this case).
The behavior for $n/V_S\gtrsim 0.2$-$0.5$ should not be
taken too seriously, since the corresponding values of $\lambda$ 
are already close to the ultraviolet
regime (see~\cite{LBseminal} for more details).

\subsubsection{Bifurcation phase and $C_b$-$C_{dS}$ phase transition}\label{subsubsec:bifurc}
\begin{figure}[t]
    \includegraphics[width=0.9\columnwidth, clip]{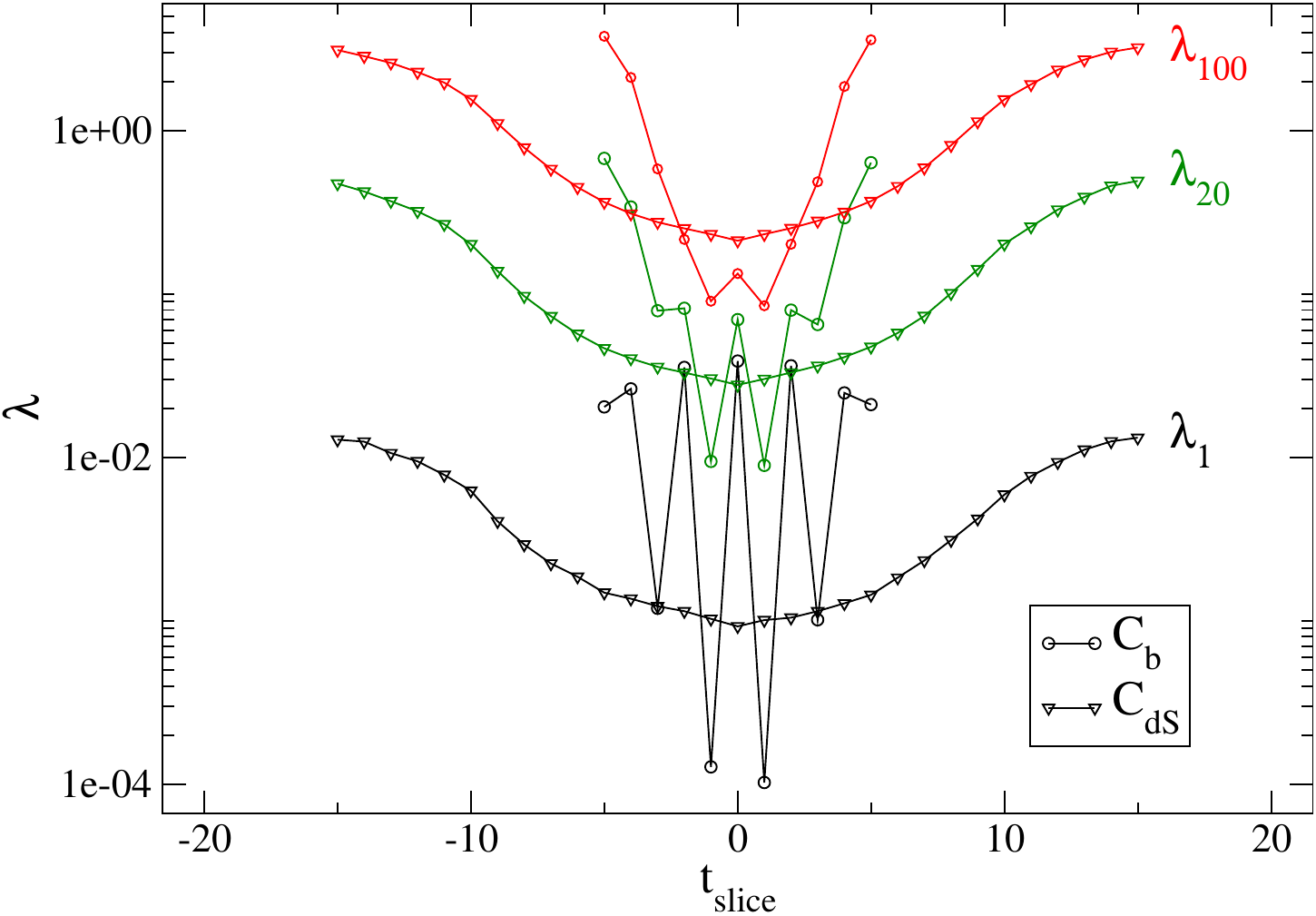}
  	\caption{Averages $\lambda_1$,$\lambda_{20}$ and $\lambda_{100}$ for configurations in $C_b$ and $C_{dS}$ phases
   and as a  function of the slice-time. The slice-time of maximal slices has been shifted to zero. See Ref.~\cite{LBseminal} for details.}
    \label{fig:aver_Cb-CdS_lam1-20-100_vs_tslice}
\end{figure}

\begin{figure}[t]
    \includegraphics[width=0.9\columnwidth, clip]{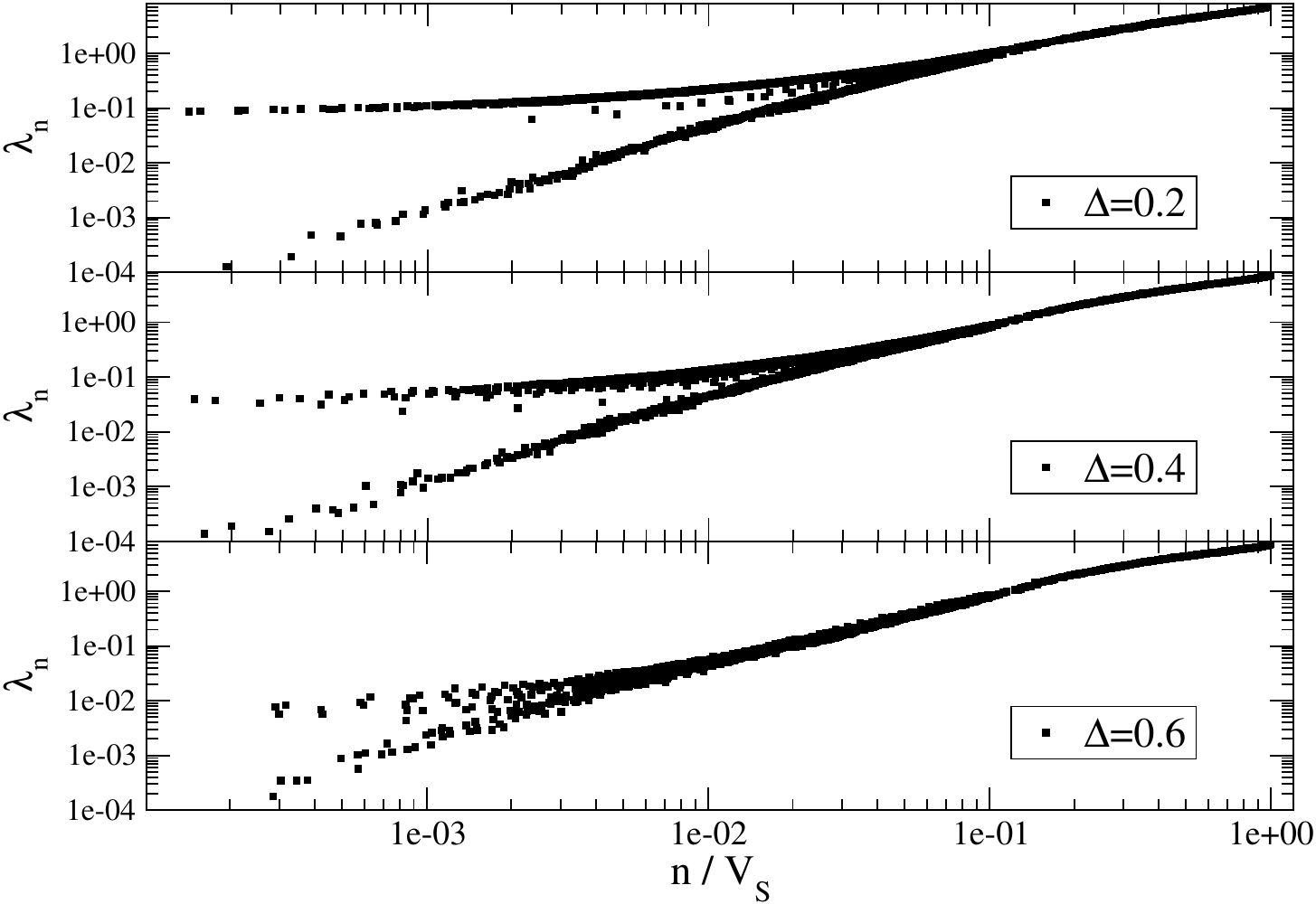}
    \caption{Scatter plot of $\lambda_n$ versus $n/V_S$ for the slices (with spatial volume $V_S > 200$) of single configurations sampled at $k_0=0.75$, and for three different values of $\Delta$ in the $C_b$ phase.See Ref.~\cite{LBseminal} for details.}
    \label{fig:lam_koV_compr}
\end{figure}

\begin{figure}[ht]
    \centering
    \includegraphics[width=\linewidth]{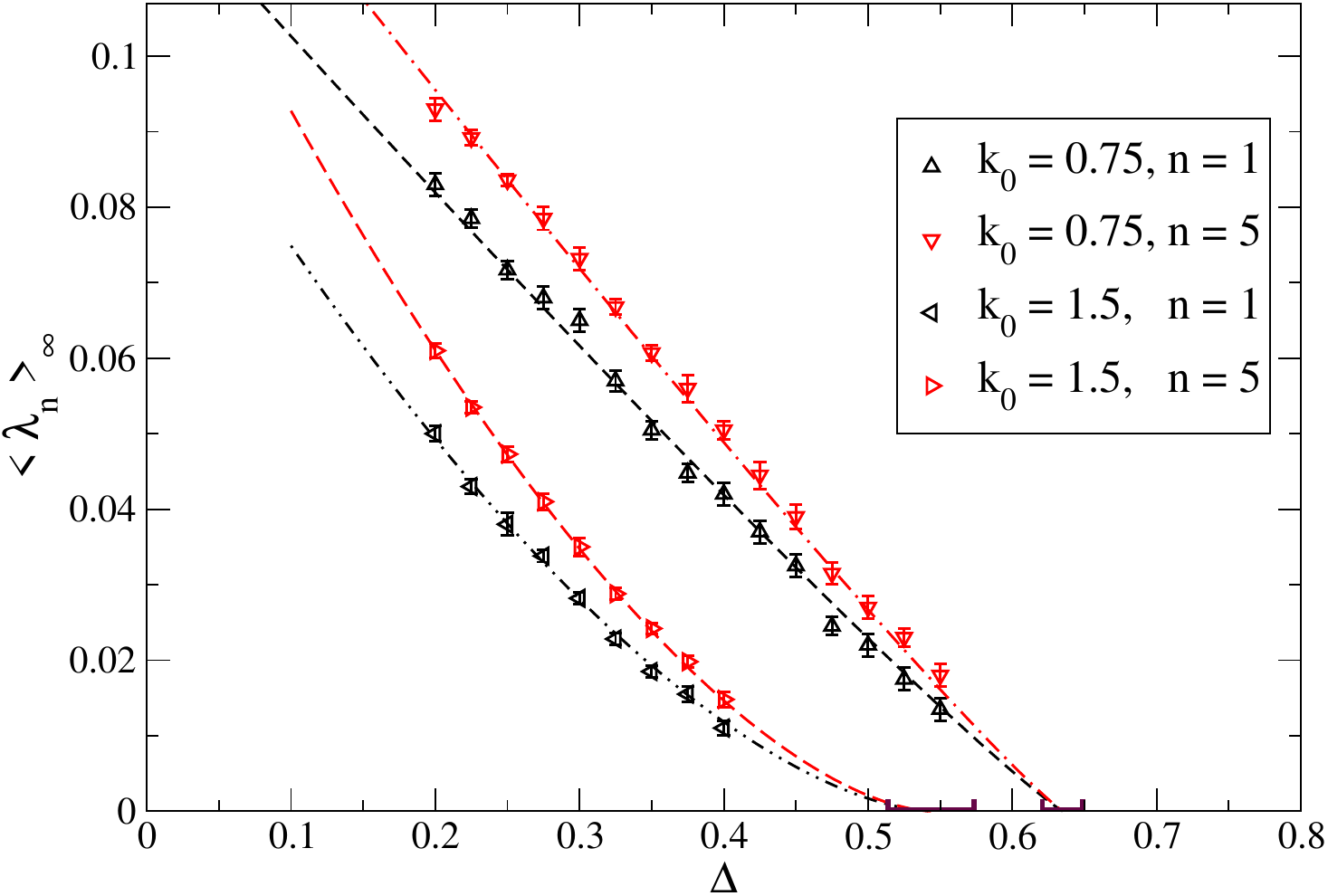}
    \caption{Critical scaling of the spectral gap and lowest order eigenvalues approaching the $C_b$-$C_{dS}$ phase transition and best fit with Eq.~\eqref{eq:critfit}.}
    \label{fig:bicritical}
\end{figure}

Here we complete the spectral characterization 
of slice geometries by considering the ensembles of configurations
in the bifurcation phase $C_b$, which exhibits the same typical
extended volume profile as the $C_{dS}$ case, but which has been
found to differ in the presence of a structure with  
vertex coordination numbers alternating in slice time between relatively small and large values. From this separation, 
$C_b$ takes the name ``bifurcation'' phase.
The methods are completely general, 
but in the following, we always discuss 
results for spatial slices with spherical topology.
From the spectral point of view, 
Figure~\ref{fig:aver_Cb-CdS_lam1-20-100_vs_tslice} shows 
a qualitative comparison between the average spectral gaps 
(and a few higher-order eigenvalues) of $C_{dS}$ and $C_b$ slices 
as a function of the slice time. 
The alternating structure in slice time is quite noticeable
in the slices from configurations in $C_b$ phase, 
while it does not appear in the ones from $C_{dS}$ phase.
Furthermore, it is interesting to notice that for 
at some scale (around $\lambda\gtrsim \lambda_{100}$ in 
Figure~\ref{fig:aver_Cb-CdS_lam1-20-100_vs_tslice}) 
how the qualitative dependence on the slice time 
seems to approach the $C_{dS}$ one. 
Since relatively large spectral gaps are associated to highly
connected geometries at large scales, as we showed for
$B$-phase slices, 

As discussed in Chapters 1 and 10 of this Section~\cite{Ambjorn_chapt1,Gizbert-Studnicki_chapt10}, 
it is apparent that $C_{dS}$ and $C_{b}$ are the only physically relevant phases. 
Indeed, the average spatial volume distribution observed for ensembles 
in the $C_{dS}$ phase with spherical slice topology is in good agreement with what is expected from
a de Sitter Universe, which corresponds to a $S^4$ geometry 
after analytical continuation to the Euclidean space~\cite{cdt_desitter}. 
Unlike $C_{dS}$, the bifurcation phase $C_b$ is characterized instead 
by the presence of two different classes of slices that alternate each other 
in slice time $t$~\cite{new_phase_chars,cdt_newhightrans}.
This bifurcation behavior of $C_b$ phase is apparent 
also from the general behavior of the aggregated spectrum of all slices as shown 
in Fig.~\ref{fig:lam_koV_compr} 
for three different values of $\Delta = 0.2, 0.4$ and $0.6$, 
at the line with constant $k_0 = 0.75$:
inside the bifurcation phase, some slices are gapped, but increasing the
value of $\Delta$ the gap disappears and the two classes of slices merge
into a common $C_{dS}$-like slice behavior.
In particular, one can compute the average 
over an ensemble of configurations
at fixed simulation parameters and the infinite volume extrapolation $\langle \lambda_n \rangle_\infty$
for each eigenvalue order $n$. 
On the left side vicinity of the critical point, 
this quantity follows a shifted power law 
\begin{align}\label{eq:critfit}
\langle \lambda_n \rangle_\infty  = A_n (\Delta - \Delta_c)^{2 \nu},
\end{align}
where $\Delta_c$ and $\nu$ are in general functions 
of $k_0$. Only the coefficients $A_n$ of Eq.~\eqref{eq:critfit} depend on the eigenvalue order $n$,
while $\nu$ appears to be the same for the lowest orders, with some tension for $n\gtrsim 10$.
For example, results of a combined fit with Eq.~\eqref{eq:critfit}, 
including specifically the orders $n = 1$ and $n = 5$, 
are shown in Figure~\ref{fig:bicritical}, and yield
$\Delta_c = 0.635(14)$, $\nu = 0.55(4)$ for 
$k_0 = 0.75$ ($\chi^2/{\rm d.o.f.} = 31/26$), and 
$\Delta_c = 0.544(36)$, $\nu = 0.82(12)$ for 
$k_0 = 1.50$ ($\chi^2/{\rm d.o.f.} = 6/14$).
Since eigenvalues at each order $n$ can be interpreted as distinct 
characteristic long-range lengths, 
the observation that they appear to scale with the same exponent $\nu$
suggests that a single length scale 
(e.g.~the spectral gap $\lambda_1$) determines the general scaling behavior 
at the large length scales. 
This is reassuring 
because it is expected to happen when continuum physics behavior,
where all quantities can be described by the scaling of
a single characteristic length $\xi$, typically in the form of a
correlation length. Therefore gap between $C_b$ and $C_{dS}$ 
phases closes as expected from a second-order phase transition,
and for both the values of $k_0$ investigated. 

\subsection{An alternative approach to the Laplacian discretization: Finite Element Methods}
\label{subsec:FEM}

The graph discretization of the LB operator is certainly useful, 
but in general, other discretizations are possible, 
as well as extensions to the space of vector or tensor fields,
as investigated in Ref.~\cite{Reitz:2022dbj}, 
where generalized spectral dimensions are defined.
Here we introduce a Finite Element discretization as 
an alternative tool that allows investigating some features 
not available in general with graph or other discretization 
in a consistent way.
This Section is based on Ref.~\cite{LBFEMseminal}, 
which should be referred to for a more detailed discussion.

\subsubsection{Weak formulation and the Finite Element Methods}
The name Finite Element Methods (FEM) covers a wide family
of approximation techniques that are applied in many fields, where complex modeling is necessary, to numerically solve very general integro-differential equations ~\cite{fem_allairebook,fem_hughesbook,fem_strangbook,fem_taylorbook,fem_babuska}. As we describe in more detail in the following, 
the general procedure consists in casting the components of the problem
under investigation into simpler and smaller parts, 
which makes the problem easier to be treated numerically.

In the context of performing spectral analysis of general manifolds
FEM consists in casting the LB eigenproblem 
\begin{equation}\label{eq:eigp}
    -\triangle f(\mathbf{x})= \lambda f(\mathbf{x}).
\end{equation}
into a weak formulation by multiplying both sides by a test function $\phi(\vec{x})$
and integrating everywhere, which results, for a boundaryless manifold $\mathcal{M}$, 
in 
\begin{equation}\label{eq:weak-eigp}
\int_\mathcal{M}\! d^dx \;\nabla \phi(\mathbf{x}) \nabla f(\mathbf{x}) = \lambda \int_\mathcal{M}\! d^dx \; \phi(\mathbf{x}) f(\mathbf{x}),
\end{equation}
where integration by parts has been performed on the left side.
The test function $\phi$ can belong to different classes of functions,
but it is customary for the LB eigenproblem to consider the Sobolev space $H^1(\mathcal{M})$,
or simply $H^1$, defined as the space of $L^2$ functions that admit weak first derivatives,
since the problem is well-posed and solutions are proven to exist.
We want to stress that, for a piecewise-linear manifold $\mathcal{M}$, 
the Sobolev space $H^1(\mathcal{M})$ involves all functions on the whole domain of
the manifold, i.e., not just the vertices or the simplexes centers, but also the interior
of flat simplexes. In the following, we call the spectrum of the Laplace--Beltrami operator
on the space of functions $H^1(\mathcal{M})$ the \emph{exact LB spectrum}. 
However, this space is infinite-dimensional and cannot be treated numerically, 
even in the weak form shown in Eq.~\eqref{eq:weak-eigp}.
One then needs to set up an approximation scheme by building a sequence
$\{\mathcal{V}_r\}_{r=0}^\infty$ of finite-dimensional subspaces of $H^1$
with increasing dimension such that in the limit one recovers 
the full space $\lim_{r\to\infty} \mathcal{V}_r \to H^1$.
Accordingly, the approximate eigenvectors on $f_n^{(r)}$ and eigenvalues $\lambda_n^{(r)}$
would converge, in the limit $r\to\infty$ to the exact LB eigenvectors and eigenvalues 
of the infinite-dimensional problem~\eqref{eq:weak-eigp} in $H^1$.
In practice, chosen a finite set of basis functions $\{\phi^{(r)}_i\}_{i=1,\dots,N_r}$
for the subspace $\mathcal{V}_r$, any function $f\in H^1$ can approximated as
\begin{equation}\label{eq:FEMfuncf}
f(\mathbf{x})=\sum_{i=1}^{N} c_i \phi_i(\mathbf{x}).
\end{equation}
With this expansion, and the basis used as test functions, 
Eq.~\eqref{eq:weak-eigp} can be rewritten in the
form of a finite-dimensional generalized eigenvalue problem:
\begin{equation}\label{eq:fem-eigp}
L \vec{c} = \lambda M \vec{c}
\end{equation}
where we have introduced the two matrices $L$ and $M$ with matrix elements:
\begin{align}\label{eq:FEM_genL}
L_{i,j} &\equiv \bigintssss_{\mathcal{M}} \!d^d \mathbf{x}\,  \vec{\nabla}\phi_i(x) \cdot \vec{\nabla}\phi_j(x), \\[5pt]\label{eq:FEM_genM}
M_{i,j} &\equiv \bigintssss_{\mathcal{M}}\!d^d \mathbf{x}\, \phi_i(x) \phi_j(x).
\end{align}
For the specific class of FEM we considered,
the procedure of building a consistent sequence of subspaces 
is called \emph{refinement},
and it allows extrapolating the result of solutions of the approximate LB eigenproblem
on a finite set of subspaces $\mathcal{V}_r$ to the infinite \emph{refinement level} 
$r$ limit, which corresponds to $H^1$. The details of this construction
and the convergence properties of the extrapolation to infinite refinement level are technical
has been left out of the following discussion, where we assume all results 
as already extrapolated, 
but the reader can find a comprehensive discussion in Ref.~\cite{LBFEMseminal}. 
The main point of the FEM discretization of the LB
operator is actually the fact that it allows one to extract, by extrapolation, 
an arbitrarily good approximation to the result of the eigenproblem 
on the full infinite-dimensional Sobolev space of functions $H^1$ 
on the piecewise-flat manifold under investigation. 
Of course, while powerful, this approach is certainly not numerically cheap.
However, we think it is useful in order to check how much results obtained 
with other discretizations of the LB operator deviate from the exact spectrum 
and eigenfunctions, as we discuss in the next Section.

\subsubsection{Numerical Results}\label{subsubsec:numresFEM}
Here we show a comparison between the few smallest eigenvalues 
of the LB operator acting on the Sobolev space of functions $H^1$ 
for CDT slices, accessed through FEM, and the results of the dual graph discretization
of the LB operator, as described in Section~\ref{subsec:lap_dg}.
In Ref.~\cite{LBFEMseminal}, a detailed discuss 
The effective dimension obtained with the two methods is investigated in 
Ref.~\cite{LBFEMseminal} for slices in two points in the phase diagram.
For example, for $(k_0,\Delta)=(0.75,0.7)$ 
the best effective dimension estimate via FEM suggests a value $d^{(FEM)}_{EFF}=2.088(18)$,
which is in contrast with the value $d^{(DG)}_{EFF}\simeq 1.6$ as found 
in Ref.~\cite{LBseminal}.
For the sake of comparison, we show here how the estimate of the 
critical index of the $C_b$-$C_{dS}$ transition along the line $k_0=0.75$ 
differs from the one computed using dual graph methods, 
which has been discussed in Section~\ref{subsubsec:bifurc}.

In order to extract results useful for the continuum limit, 
the following three limiting procedures should be performed 
\begin{enumerate}
    \item for each simplicial manifold $\mathcal{M}$, one has to extrapolate 
    the individual FEM eigenvalues to ``infinite refinement level'' 
    $\lambda_n^{(r)}[\mathcal{M}] \xrightarrow{r\to \infty} \lambda_n^{(\infty)}[\mathcal{M}]$, in order to obtain an accurate enough approximation of the spectrum of the exact LB differential operator on $\mathcal{M}$ (see Ref.~\cite{LBFEMseminal} for details
    about this step);
    \item for each ensemble of configurations at specific values of the parameters, 
        one should first perform an average of the eigenvalues at each specific order $n$ 
        and then take the thermodynamic limit (i.e., infinite volumes in lattice units) 
        \mbox{${\langle \lambda_n \rangle}_V \equiv \frac{1}{\lvert \mathcal{C}_{V} \rvert} \sum_{\mathcal{M}\in \mathcal{C}_V} \lambda_n^{(\infty)}[\mathcal{M}] \xrightarrow{V\to \infty}  {\langle \lambda_n \rangle}_\infty$} by considering the spectra of ensembles
        $\mathcal{C}_{V}$ with increasing volumes;
    \item finally, one can study the critical scaling of the twice-extrapolated eigenvalues ${\langle \lambda_n \rangle}_\infty {(k_0,\Delta)}$ observed as the phase transition is approached and match it with the shifted power law expressed by Eq.~\eqref{eq:critfit}.
\end{enumerate}

In the rest of this Section, we considered only some points along the line $k_0=0.75$, 
and the critical scaling of the first ten orders of eigenvalues. 
First, we have to possibly extrapolate to infinite refinement level, 
and then we can perform the thermodynamic limit ($V_S \to \infty$) order by order.
For this second step, in general we do not have a precise expectation 
on the large-scale behavior for the eigenvalues of gapped slices, 
but we followed~\cite{LBrunning} and used the simplest form compatible with data, 
that is, a quadratic polynomial in $1/V_S$:
\begin{equation}\label{eq:thermlim}
\langle \lambda_n \rangle_V = \langle \lambda_n \rangle_{\infty} + \frac{A_n}{V_S} + \frac{B_n}{V_S^2}.
\end{equation}
The thermodynamical limit consists then of fitting data with Eq.~\eqref{eq:thermlim} 
in order to extract the possibly non-zero constant terms $\langle \lambda_n \rangle_{\infty}$,
order by order in $n$.
\begin{figure}[ht]
	\centering
	\includegraphics[width=\linewidth,trim={0.72cm 0.27cm 0.70cm 1.38cm},clip]{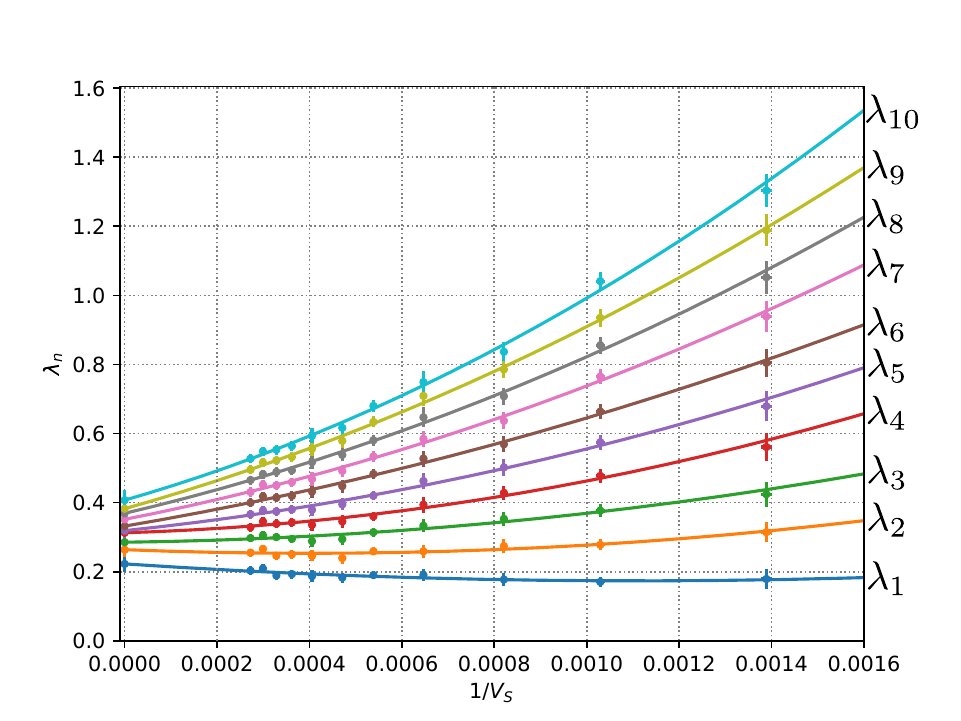}
	\caption{First ten eigenvalue orders of spatial slices (with $V_S>500$) 
    vs $1/V_S$, with extrapolation to the thermodynamic limit ($V_S\rightarrow\infty$) 
in $C_b$ phase. Phase space point $(k_0,\Delta)=(0.75,0.7)$, total spatial volume $V_{S,tot}=80k$.}
    \label{fig:thermlim}
\end{figure}
For every point in the phase diagram and every order $n$ taken into account, 
the thermodynamic limit extrapolations yielded $\chi^2/dof<1$. 
For illustration purposes, Figure~\ref{fig:thermlim} displays the extrapolation 
to the thermodynamic limit for the first ten eigenvalue orders in the phase space point 
$k_0=0.75, \Delta=0.575$.
\begin{figure}[ht]
	\centering
	\includegraphics[width=\linewidth,trim={0.72cm 0.36cm 1.13cm 1.45cm},clip]{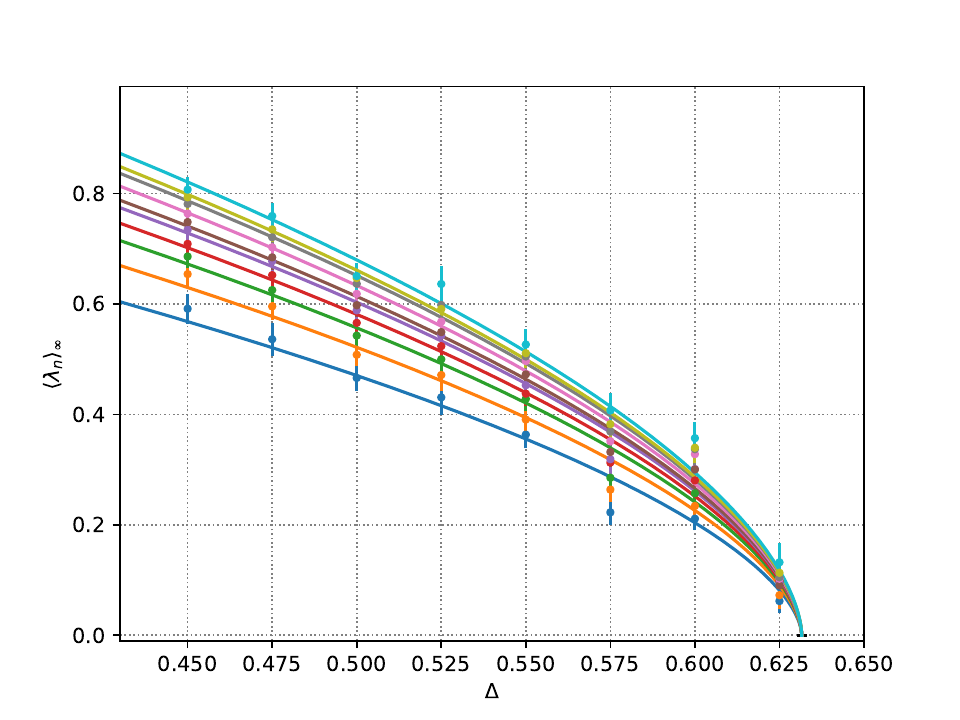}
	\caption{Critical behavior of the first ten eigenvalue orders along the 
    line at fixed $k_0=0.75$ and varying $\Delta$, with best-fit curves of the form shown 
in Equation~\eqref{eq:critfit} with common $\nu$ and $\Delta_{crit}$.
    Curves of increasing eigenvalue order are shown from below to above in the plot.}
    \label{fig:critscaling}
\end{figure}
We then analyzed the first ten eigenvalue orders by fitting our data with 
the shifted power law in Eq.~\eqref{eq:critfit}, 
imposing a common critical index $\nu$ and 
critical point $\Delta_{crit}$ for every order $n$, 
as observed also for the dual graph case
discussed in Section~\ref{subsubsec:bifurc}. 
We used data coming from eight phase space points with $\Delta$ ranging 
from $0.45$ to $0.625$: we chose not to go too deep inside $C_b$ phase because of 
the influence of the expected sub-dominant terms of the critical scaling 
and excluded them by checking the stability of our estimate of $\nu$ under 
the removal of the points with lower $\Delta$ parameter. 
We obtained, as best-fit parameters $\nu=0.293(10)$ (recall that $\nu^{(DG)}=0.55(4)$ in Section~\ref{subsubsec:bifurc} at the same value of $k_0$)
and $\Delta_{crit}=0.6316(15)$, with reduced chi-squared $\frac{\chi^2}{\text{dof}} = \frac{67}{68}\simeq 1$. 
Data and best-fit curves for each order up to $n_{\text{max}}=10$ are displayed in Figure~\ref{fig:critscaling}.

In conclusion,
it is apparent that data are still qualitatively compatible with 
the critical scaling from dual graph spectra discussed 
in Section~\ref{subsubsec:bifurc} 
and based on~\cite{LBrunning}, but, 
while the location of the transition line agrees with those results, 
the values found for the critical indexes appear significantly
different than the previous estimates. 
Such a difference might be important in assessing the quality of
estimates based on graph discretization,
in particular, if critical indexes of different observables have 
to be compared to find a physical continuum limit in the phase diagram.

\section{Gauge fields on fluctuating geometries}
\label{sec:gf}

The study of Dynamical Triangulations in the presence of additional quantum fields has been considered frequently in the literature. This is quite natural: if CDT will reveal a successful way to quantize QG, it should eventually be considered in connection with other interacting fields; on the other hand, the presence of other fields may by itself enter the analysis of the renormalization group flow and the search for the continuum limit.

Abelian and non-Abelian gauge fields are a notable case, given their role in the formulation of the standard model. A first important aspect concerns the way the local gauge symmetry and the gauge fields are implemented in the context of Dynamical Triangulations. Gauge fields in a discrete setting, e.g., a hypercubic lattice, are usually described~\cite{Wilson:1974sk} in terms of elementary parallel transporters living on lattice links (gauge link variables), which permit to {\em translate} the local frame choice for the internal symmetry group from one lattice site to the other. Therefore, local gauge transformations act on lattice sites, which is the same place where matter fields live.

In the context of Dynamical Triangulations, it is most natural instead to make the choice for the internal symmetry consistent with that for the Lorentz symmetry, i.e. to associate gauge transformations to the simplices composing the triangulations. In this way, the elementary parallel transports are associated with the links of the dual graph, which connect adiacent simplices and are dual to the hypersurfaces separating them.

Once the nature of gauge transformations and gauge fields on the triangulation has been clarified, the first task is to write down a properly discretized action for the composite gravity-gauge system. In the following, $\YMconfs$ will stand for the space of all 
possible gauge field configurations on the triangulation $\mcTau$, where $G$ is the gauge group. 
The action can be formally
decomposed, in the minimal coupling paradigm, 
as $S[\Phi,\mcTau] = S_{CDT}[\mcTau] + S_{\text{YM}}[\Phi;\mcTau]$, 
with $\Phi \in \YMconfs$ and where $S_{\text{YM}}$ represents the minimally 
coupled gauge action in the gravity background $\mcTau$. The specific form of $S_{\text{YM}}$, which correctly represents Yang-Mills theory in the continuum limit, can be easily built taking as a guide what is usually done on a standard hypercubic lattice~\cite{Wilson:1974sk}, as we discuss in more detail in the following.

\subsection{Yang-Mills action coupled to Dynamical Triangulations}

Continuum Yang-Mills (YM) theories are described in terms of the gauge field
$A_{\mu}=A^{a}_{\mu} T^{a}$, where $T^{a}$ are the generators of the Lie algebra ($a\in\{1,\dots,N^2-1\}$ for $SU(N)$, normalized as
  ${\rm Tr}(T_a T_b) = \delta_{ab} /2$),
while the continuum action in flat Euclidean space reads 
\begin{equation}\label{eq:un-YMact_flat}
    S_{\text{YM}} = \frac{1}{4} \bigintssss\limits_{\mathcal{M}} \! d^d x \; F^{a}_{\mu \nu} F^{a \mu \nu}
,
\end{equation}
where $F^{a}_{\mu \nu}=\partial_\mu A^{a}_{\nu} - \partial_\nu A^{a}_{\mu} + g f^{abc} A^{b}_{\mu} A^{c}_{\nu}$.  A minimal coupling to gravity (described in terms of the Einstein-Hilbert action) leads to the following modification:
\begin{equation}\label{eq:un-YMEHact_flat}
    S_{\text{YM + EH}} = 
\bigintssss_{\mathcal{M}} \! d^d x \sqrt{-g} \; 
\left[ \frac{1}{4} F^{a}_{\mu \nu} F^{a \mu \nu} + (R - 2 \Lambda) \right],
\end{equation}
where $g$ in $\sqrt{-g}$ stands now for the determinant of the metric tensor,
$R$ is the scalar curvature and $\Lambda$ the cosmological constant.

In flat space-time, YM theories are usually discretized 
on a hyper-cubic lattice in terms of link variables $U_\mu(n)$, 
representing the elementary parallel transporters from lattice site
$n$ to lattice site $n + \hat \mu$ and taking values
in the gauge group $G$. Local gauge transformations $g(n)$ act on lattice sites,
and gauge link variables transform as $U_\mu(n) \to g(n) U_\mu(n) g^{-1}(n + \hat \mu)$.
The standard and simplest gauge invariant discretization 
of the action is given in terms of the so-called {\em plaquette}
operator (plaquette or Wilson action):
\begin{equation}\label{eq:un-YMact}
S_{\text{YM}} \equiv - \frac{2N}{g^2} \sum\limits_{\Box} \Big[ \frac{1}{N} Re Tr \Pi_\Box - 1 \Big] \, ,
\end{equation}
where $\Pi_\Box$ stands for the oriented product of gauge link variables around 
an elementary plaquette $\Box$, and the sum extends over 
all possible plaquettes\footnote{For the Abelian gauge group $U(1)$, 
the factor $2N/g^2$ is substituted by $1/g^2$.}. The fact that Eq.~\eqref{eq:un-YMact}
correctly reproduces continuum YM in the na\"ive continuum limit follows from the
second order expansion of the
correspondence between plaquettes and continuum field strengths, 
$\Pi_{\mu\nu} \simeq \exp\left(i g a^2 F_{\mu\nu}\right)$, where $\mu\nu$ fixes the
plaquette orientations ad $a$ is the lattice spacing, which in turn derives from the 
correspondence $U_\mu \simeq \exp\left(i g a A_\mu\right)$.

On a triangulated, curved space-time instead, as already explained above, we consider
a formulation where elementary parallel transporters are associated with dual links connecting
pairs of adiacent simplices. In the following, such 
variables will be indicated as $U_\mu(s)$, where $s$ stands for a particular
simplex and $\mu$ is the dual link direction, or simply as $U_l$, where
$l$ is the dual link joining two simplices $s$ and $s'$; 
under a local gauge transformation $g(s)$, $U_l \to g(s) U_l g^{-1}(s')$.
All dual links connect ideally
the centers of the simplices, so they all have equal lengths, except when 
simplices are anisotropic, as usual in 4D CDT.
Gauge invariant objects are associated
with traces of closed loops over the dual graph and, in analogy with 
the standard formulation, we will call \emph{plaquette}
the most elementary closed loop, as well as the gauge invariant operator associated with it.
 A plaquette encloses an elementary 
2D surface of the dual graph, which is dual to a 
$(d-2)$ simplex of the original triangulation, i.e.~a \emph{bone} in the Regge terminology~\cite{regge}:
this is the geometrical object where space-time curvature resides,
now it becomes the object where the \emph{gauge curvature} lives as well.

A plaquette $\Pi_b$ corresponds to the ordered product 
of $n_b$ dual link variables going around bone $b$,
where $n_b$ is the coordination number of the bone.
The area enclosed by an elementary plaquette is 
$A n_b$, where $A$ is the elementary unit of area 
(e.g., one third of the simplex area in two dimensions), so that the correspondence between plaquettes 
and continuum field strengths reads:
\begin{equation}
\Pi_b \simeq \exp\left(i g n_b A F_{\mu\nu}\right)
\end{equation}
where $\mu$ and $\nu$ define the plane orthogonal to the bone.
The second order expansion of the trace of the plaquette 
returns $F_{\mu\nu}^2$ also in this case, however
accompanied by a factor $n_b^2$ which, contrary to what happens on a lattice
with a fixed and homogeneous geometry, is a dynamical variable 
that must be properly taken care of.

In order to understand how, let us consider that, 
in the continuum action, $F_{\mu\nu}^2$ gets multiplied by
$d^d x \sqrt{-g}$ factor, which measures the physical volume. 
If we want to count volume while summing over bones, we have to take
into account the volume pertaining to each bone, which is proportional 
(times a constant geometrical factor for isotropic simplices) 
to the number of simplices sharing the bone, i.e.~to the coordination
number $n_b$. From this reasoning it is clear that, since the second order
expansion of the plaquette returns $F_{\mu\nu}^2$ with a $n_b^2$ factor attached, one needs 
to divide by $n_b$ the contribution from each plaquette and then sum over all plaquettes.
The form of the plaquette action over 
a triangulation $\mcTau$ then reads: 
\begin{equation}\label{eq:un-YMact_cdt}
    S_{\text{YM}} \equiv - \beta\!\!\!\! \sum\limits_{b \in \mcTau^{(d-2)}}\!\! \frac{\widetilde{\Pi}_b}{n_b},
\end{equation}
where we adopted the shorthand 
\mbox{$\widetilde{\Pi}_b\equiv \lbrack \frac{1}{N} Re Tr \Pi_b - 1 \rbrack$}, 
$\mcTau^{(d-2)}$ is the set of all bones of the triangulation $\mcTau$, while 
$\beta$ denotes, as usual, the inverse gauge coupling proportional to $1/g^2$.
This expression is valid for any dimension $d$, however the exact definition of 
$\beta$ includes geometrical factors (like the elementary area $A$ and the 
simplex volume) which depend on $d$.
\\

\begin{figure}[t!]
\centering
\includegraphics[width=0.8\textwidth]{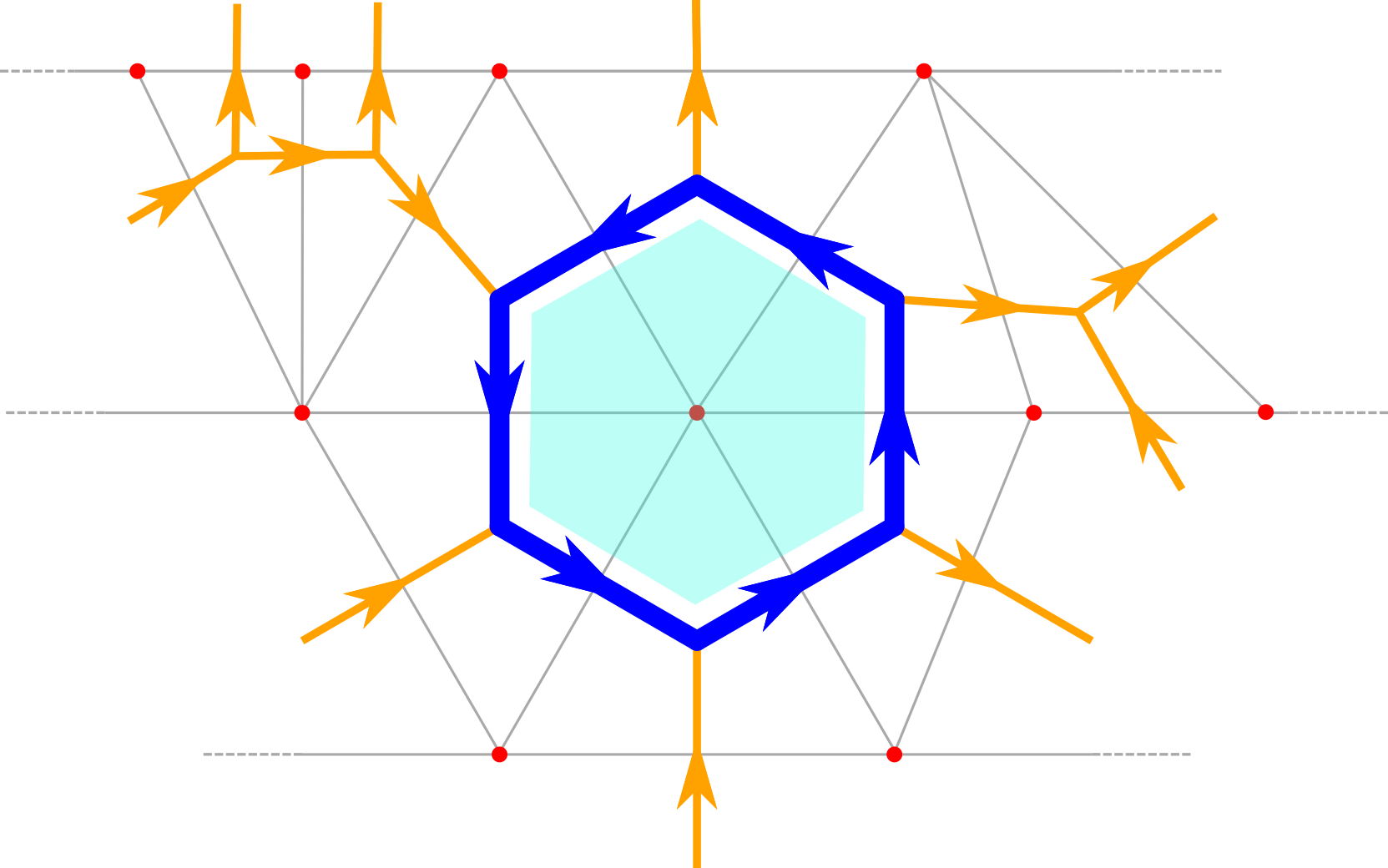}
\caption{Sketch of a plaquette operator on a typical 2D CDT triangulation. 
We have considered a plaquette built around a vertex with $n_b = 6$, corresponding to zero 
curvature. The direction of gauge links (apart from those belonging to the plaquette)
has been drawn according to the convention adopted in
the text, i.e.~from bottom to up and from left to right.
All triangles in this and the following figures 
should be considered as equilateral, even if they are necessarily drawn differently.}
\label{fig:plaquette_draw}
\end{figure}

In the following, we will consider an explicit realization of the above formulation
in two space-time dimensions. In this case the gravity sector becomes particularly simple:
the curvature term depends only on the global topology of the manifold (Gauss-Bonnet theorem), which is 
usually kept fixed, so it can be ignored and the cosmological constant term
is the only non-trivial coupling.
In particular, the pure-gravity contribution for CDT action in Euclidean space-time reads~\cite{cdt_review12}:
\begin{equation}\label{eq:mincoup-CDT2Daction}
    S_{\text{CDT}}^{(\text{2D})}[\mcTau]= \lambda N_2[\mcTau],
\end{equation}
where $N_2[\mcTau]$ is the number of simplices (triangles) in $\mcTau$ 
and $\lambda$ is the only coupling, related to
the cosmological constant. When considering the integral over all possible triangulations weighted 
by $\exp( - S_{\text{CDT}}^{(\text{2D})}[\mcTau])$, a critical behavior is observed  
as $\lambda \to \lambda_c = \log 2$; in particular, both the average 
total volume and the correlation length for foliation volumes 
diverge as $\lambda \to \lambda_c$ from above.

In Fig.~\ref{fig:plaquette_draw} we show an explicit realization of 
a configuration of gauge link variables $U_\mu(s)$ living on a 2D causal triangulation,
and of a plaquette built with them.
In this case $\mu$ is either spatial or temporal, with positive orientations 
taken rightward in space and upward in time; each simplex is associated
with one temporal link (either ingoing or outgoing)
and two spatial links (one ingoing and one outgoing).
In the particular example of plaquette shown in 
the figure, we have $n_b = 6$, which corresponds to a locally flat space-time
(while $n_b > 6$  and $n_b < 6$ correspond, respectively, 
to negative and positive local curvature).

\subsection{Numerical simulations of CDT coupled to YM theories}
\label{subsec:algo}

A Monte-Carlo approach to the computation of path-integral averages requires to sample the possible 
triangulation+gauge field configurations according to 
their weight, $\exp(-S_{\text{CDT}}^{(\text{2D})} - S_{\text{YM}})$, 
with $S_{\text{YM}}$ defined in Eq.~\eqref{eq:un-YMact_cdt}. In order to that,
one has to devise a set of Markov chain moves
which guarantee detailed balance, ergodicity and aperiodicity.
Since the action is local, the natural choice is to look for a set
of local moves, which change the triangulation and/or the gauge configuration only locally.
Moreover, it seems convenient to think of moves which change alternatively either
the gauge field or the triangulation, so that one can try to implement standard
algorithms already used either for lattice gauge theories or for CDT. 

However, while this idea works smoothly for updating gauge configurations, moves regarding the triangulation
present some additional difficulties. Indeed, any change in the space-time geometry modifies the gauge connection
as well: some simplices are added/destroyed or differently glued with the rest of the triangulation, implying 
a modification of the dual graph and of the gauge link variables living on it. 
Therefore, one can refer to standard CDT moves (for a review, see Ref.~\cite{cdt_review12}), however a proper modification 
must be devised, involving a certain number of gauge links which are added/destroyed or differently connected 
with the rest of the gauge configuration: as the number of space-time dimensions increases, the number of link variables
involved in each CDT move increases, making the task less and less trivial. Partial local gauge fixing can help
reducing the number of non-trivial link variables involved, and in fact permits to construct a viable algorithm
in two dimensions, which is the case more deeply discussed in the following.

\subsubsection{Algorithm for 2D CDT}

The algorithm usually adopted in pure gravity CDT simulations is a local 
Metropolis--Hastings\footnote{It is interesting to notice that Dynamical Triangulations is one of the cases where
the Hastings variant of the original Metropolis algorithm is actually needed. Indeed, since the move modifies
the space of stochastic variables, selection probabilities for the move and for its inverse in general differ, so that
their ratio contributes to the acceptance step.}
algorithm~\cite{metro,hastings}, based on a set of moves
that need to preserve the causal structure of the triangulation, i.e.~that change
the triangulation without spoiling its foliation; 
a proper set is provided by the so-called Alexander moves~\cite{alexander,cdt_2dmoves}.
In two dimensions, the set consists of just three moves, usually denoted as
$(2,2)$, $(2,4)$ and $(4,2)$, to specify the number of simplices involved before and after the move.
In the following, we provide a brief description of such moves and discuss how they 
must be modified in order to take into account the corresponding modification of the gauge configuration.
A more detailed discussion and a proof of detailed balance for the modified moves
can be found in Ref.~\cite{cdtgauge_pisa1}. 
However, before doing that, we describe the 
Markov chain steps implemented to modify the 
gauge field configuration at fixed triangulation, 
which is a standard extension of the heat-bath 
algorithms usually adopted in lattice gauge theories.
\\

\begin{figure}[t!]
\centering
\includegraphics[width=0.9\textwidth]{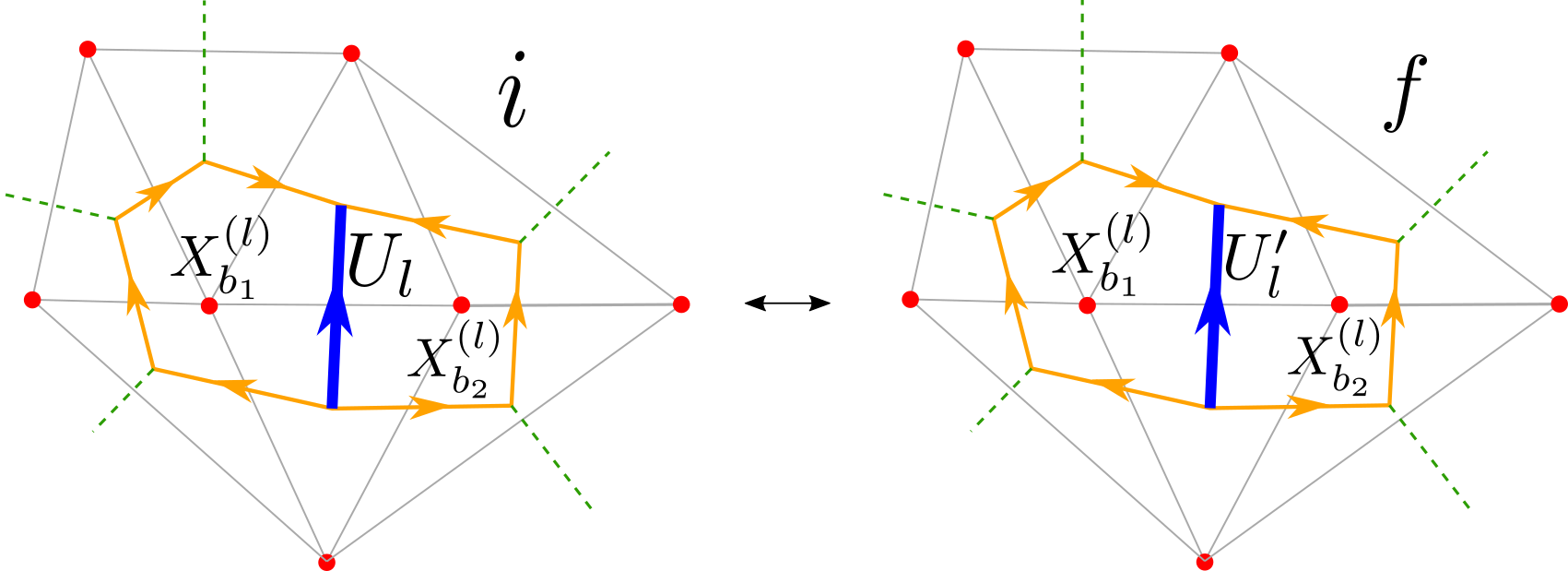}
\caption{Sketch of a typical pure gauge move on a two-dimensional causal triangulation, where a single link variable is changed, living 
on the dual link $l$ (in blue). Here and in the following, the left and right figures indicate respectively the initial ($i$) state and the final ($f$) one.
    The orientation of the products of gauge links making up the two staples
$X^{(l)}_{b_1}$ and $X^{(l)}_{b_2}$ is indicated by the arrows (orange paths).}
\label{fig:movegauge_draw}
\end{figure}

\noindent 
{\bf Pure gauge move - }
This is similar to what generally implemented in standard lattice gauge theories,
and is based on the probability distribution for a given 
gauge link variable $U_l$ ($l$ is a link of the dual lattice),
which stems from Eq.~\eqref{eq:un-YMact_cdt} for fixed values of the 
other gauge link variables:
\beq
\label{eq:link_distribution}
P(U_l) d U_l \propto 
d U_l \exp \left( \frac{\beta}{N} \textrm{Re Tr} 
\left[ U_l F_l^{\dagger} \right] \right) .
\eeq
In Eq.~\eqref{eq:link_distribution} $d U_l$ stands for the gauge invariant Haar measure over the gauge group,
while $F_l$ is the so-called local \emph{force} acting on that link, 
$F_l \equiv \sum\limits_{b \ni l} (X_b^{(l)}/n_b)$
where $X_b^{(l)}$ is the \emph{staple} 
going around the bone $b$,
i.e. the ordered product of the 
other links that form one of the plaquettes containing $l$ (in particular
that going around bone $b$).
In general, one can identify 2 such bones in two dimensions, 3 in three dimensions, 4 in four dimensions and so on;
in Figure~\ref{fig:movegauge_draw} we sketch such construction for the two-dimensional case, where the 
convention chosen for the direction of multiplication of the gauge links around the staple is also clarified.
We notice that, once the force $F_l$ has been defined, the probability distribution in Eq.~\eqref{eq:link_distribution} is identical 
to what found in standard lattice gauge theories.

The local algorithm then proceeds by selecting randomly a dual link $l$ randomly
and by updating the corresponding variable $U_l$. This can be done 
by a standard heat-bath algorithm~\cite{hb_creutz,hb_kennedy-pendleton}, i.e.~by drawing a new gauge link
variable $U_l^\prime$, in place of $U_l$, according to the distribution in Eq.~\eqref{eq:link_distribution};
the Cabibbo-Marinari algorithm~\cite{Cabibbo:1982zn} can be also easily implemented for $N > 2$, and microcanonical 
(over-relaxation) steps can be alternated to improve efficiency.
\\

\begin{figure}[t!]
\centering
\includegraphics[width=0.8\textwidth]{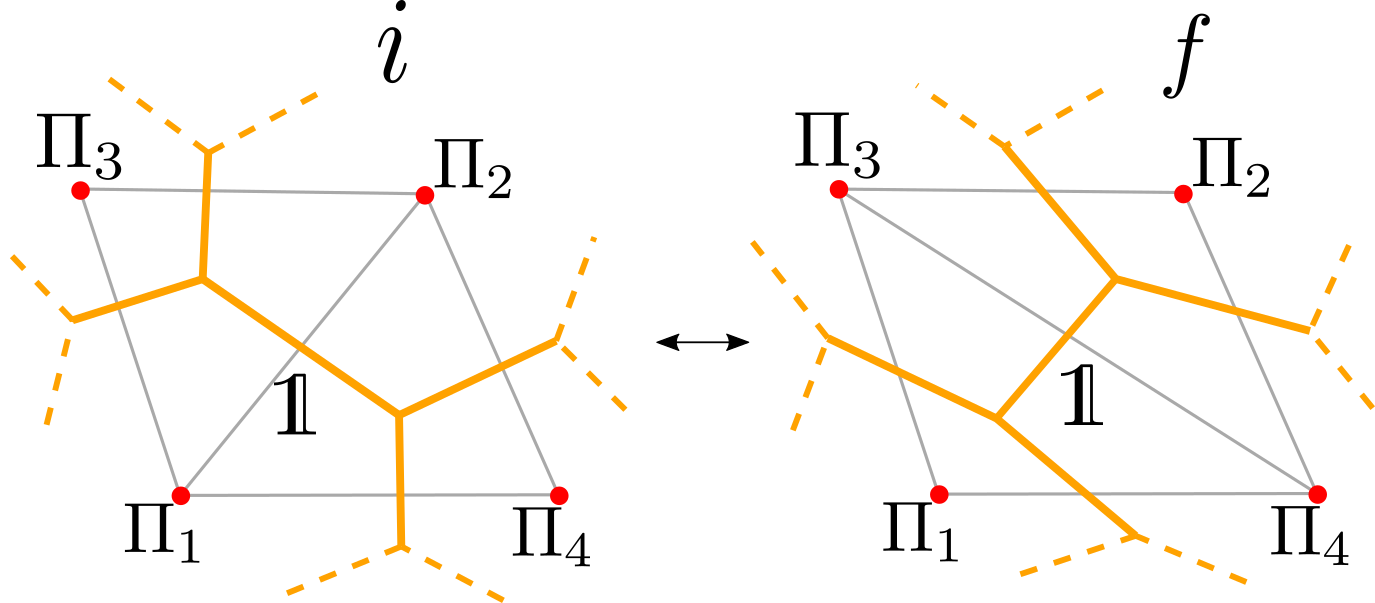}
\caption{Sketch of a $(2,2)$ move. The link dual to the flipped edge is gauge fixed to the identity before the move, so that the 4 involved plaquettes 
keep the same value after the move. However, their contribution to the action does changes, because of the modification in the coordination number of the four involved 
vertices.}
\label{fig:move22_draw}
\end{figure}

\noindent
{\bf Triangulation move $(2,2)$ - }
This move consists in flipping a time-like edge of the triangulation, hence the link dual to it, as sketched
in Figure~\ref{fig:move22_draw}. The total number of simplices is left unchanged,
so the pure gravity part of the action is untouched, however the move changes 
the space $\YMconfs$ of the gauge configurations. In particular, the coordination
number $n_1$, $n_2$, $n_3$ and $n_4$ of the four plaquettes appearing in the figure is 
changed by one unit (plus or minus), and the gauge variable living on the flipped link
enters such plaquettes in a different way. 

An easy way to proceed is to exploit gauge invariance and consider that 
the Markov move can be viewed as acting from any of the configurations which are 
gauge equivalent to the starting one, to any of the configurations which 
are gauge equivalent to the final one, i.e.~as a move between two different gauge equivalence classes (orbits). Therefore, by a proper gauge transformation,
one can always choose the starting and final configurations such that the gauge
variable living on the flipped link is gauge fixed to the identity in both cases: in practice, that means 
applying a partial gauge fixing before the move is performed. In this way, the four 
plaquettes involved in the move, $\Pi_{1\leq i \leq 4}$, are left unchanged, even if the gauge action
changes anyway, because of the modification in the coordination numbers, however this
is easily taken into account by the Metropolis-Hastings acceptance step.
\\

\begin{figure}[t!]
\centering
\includegraphics[width=0.8\textwidth]{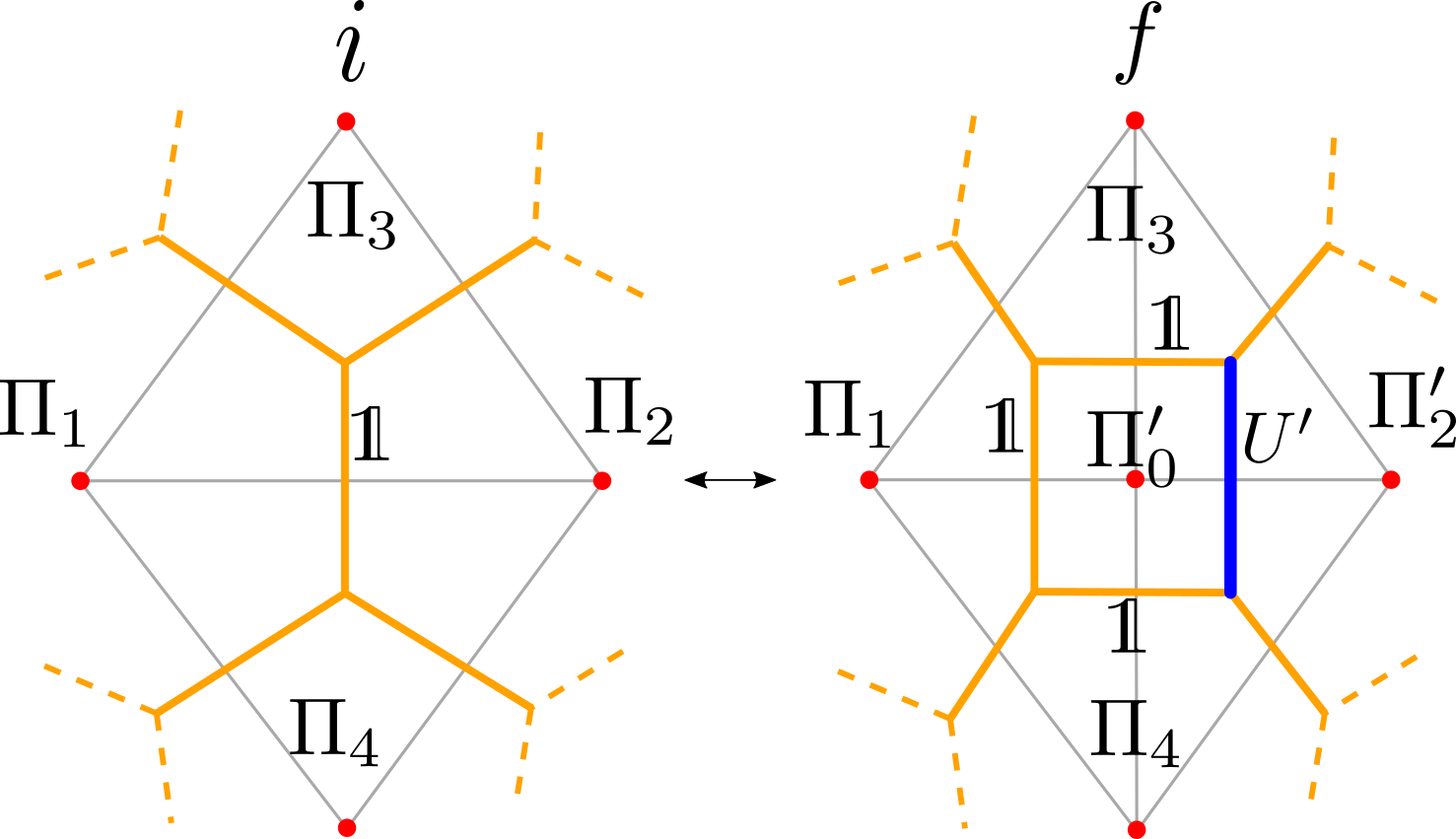}
\caption{Sketch of a $(2,4)$ move and of its inverse. Gauge fixing can be used to simplify the move only up to a certain extent. In particular, there is a starting dual link which is transformed into a plaquette of length four: while the starting link can fixed to the identity, the plaquette cannot. 
After gauge fixing, the link variable $U^\prime$ contains the non-trivial information about the new plaquette.}
\label{fig:move2442_draw}
\end{figure}

\noindent
{\bf Triangulation $(2,4)$-$(4,2)$ moves - }
These moves, which are the inverse of each other, are sketched in Figure~\ref{fig:move2442_draw}. 
They are more involved than the $(2,2)$ one.
They change the volume of the triangulation, creating
or eliminating two simplices and a vertex of coordination four,
so that the pure gravity action changes by 
$\Delta S^{(2,4)}_{CDT} = - \Delta S^{(4,2)}_{CDT} = 2 \lambda$,
and the gauge configuration is modified substantially, by 
the introduction or the removal of a central plaquette of length four,
which is indicated as $\Pi_0^\prime$ in Figure~\ref{fig:move2442_draw};
at the same time, the length and composition of plaquettes $\Pi_{2\leq i \leq 4}$
is changed as well. 

The main new difficulty with respect to the $(2,2)$ move is that, since plaquettes are gauge invariant objects,
the appearance of the new plaquette cannot be masked completely by gauge fixing.
Indeed, gauge transformations can set to the identity $\mathds{1}$ at most 
three out of four links of the central plaquette $\Pi_0^\prime$, shifting all 
the physical content in the remaining one. Therefore, during the move, one has to draw at least one 
new gauge variable (or destroy it) with a non-trivial distribution, and this must be done 
properly, i.e.~respecting detailed balance.

Let us consider for instance the $(2,4)$ move. 
A necessary condition to implement detailed balance is that we can clearly identify pairs
of states going into each other under the action of the move and of its inverse.
Therefore, while the starting link dual to the two simplices involved in the move 
can be safely gauge fixed to the identity, the new plaquette, hence the new link $U'$ in Figure~\ref{fig:move2442_draw},  
cannot, since in the corresponding inverse $(4,2)$ move, in which it is destroyed,
it will have in general a non-trivial starting value, drawn from the equilibrium distribution
of gauge fields in the target triangulation.

Therefore, the $(2,4)$ involves the extraction of a new 
link variable $U^\prime$, which then also modifies the 
value of the plaquette $\Pi_2^\prime$. That can be done in 
various ways, with the condition that detailed balance be eventually
satisfied by the final Metropolis-Hastings acceptance step. 
However, the most natural choice, which simplifies the final step
and improves acceptance, is to draw $U'$ from the heat-bath distribution 
in the target configuration, i.e.~that corresponding to the force 
$F = \mathds{1}/4 + X_2/n_2^\prime$. For more details, we refer the reader 
to Ref.~\cite{cdtgauge_pisa1}.

\subsubsection{Perspectives for higher dimensions}

In principle, the approach followed for the two-dimensional case can be extended to higher dimensions.
However, the discussion reported above for the $(2,4)$ and $(4,2)$  moves clarifies what kind of difficulties may emerge.
Let us consider, for example, the analogous $(2,6)$ and $(6,2)$ moves for three-dimensional CDT coupled to gauge fields.

In this case, two adjacent tetrahedra sharing a triangle are transformed into six tetrahedra, or viceversa, by splitting the common triangle into
three triangles. From the point of view of the dual graph, the link connecting the starting tetrahedra gets transformed into a triangular prism,
made up 9 link variables and 5 new plaquettes, two of length 3 and three of length 4. The information contained in this new geometrical 
structure can be only partially reduced by gauge fixing. In particular, gauge fixing along a maximal three can fix to the identity only
5 of the 9 links of the prism. That means that, in order to perform the move, one has to draw 4 new link variables at the same time, and take into 
account, for the final acceptance step, the modification of more than 10 plaquettes.

The above reasoning can be easily generalized to the case of generic space-time dimension $D$ for the analogous $(2,2D)$ move and its inverse.
The number of new gauge links to be created after the direct move is $2 D + 2 D (D-1)/2 = D^2 $; the number of involved dual
lattice sites usable for gauge fixing is $2 D $, however they lie on a closed hyper-surface, so the number of gauge links 
which can be actually gauge fixed to the identity is $2D - 1$. Finally, one is left with $(D-1)^2 $ new link variables to be drawn
at the same time.

This simple counting illustrates the technical difficulties emerging in higher dimensions, also related to the need for having a 
reasonable acceptance rate. As a consequence, while work is still in progress along this direction, an extension of the algorithms
developed in Ref.~\cite{cdtgauge_pisa1} to 3D and 4D CDT has not been finalized yet.
\begin{figure}[t!]
\centering
\includegraphics[width=0.9\textwidth]{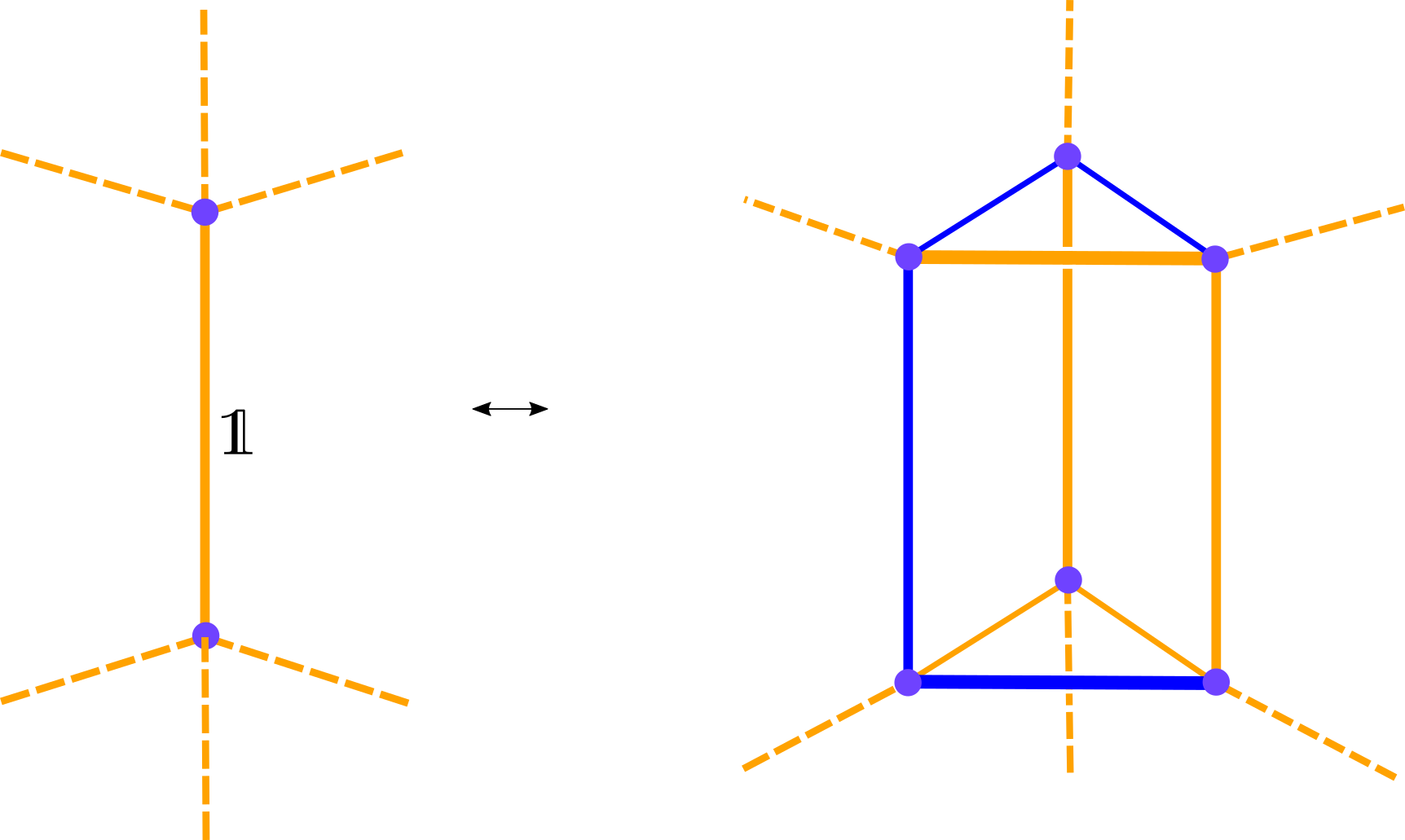}
\caption{Sketch of a possible implementation of the $(2,6)$ and $(6,2)$ moves on a three-dimensional causal triangulation, where 4 dynamical link variables (in blue) remain dynamical after gauge fixing; orange links show links gauge fixed to identity.}
\label{fig:movegauge_2662_draw}
\end{figure}

\subsection{Gauge Fields on Dynamical Triangulations at work}\label{subsec:numresYM}

In this Section we discuss some numerical results from simulations 
of 2D CDT coupled to Yang-Mills theories, considering both the
$U(1)$ and $SU(2)$ gauge group. A detailed account of such results can be found in Ref.~\cite{cdtgauge_pisa1},
here we only focus on a few aspects, which could be of particular interest when 
algorithms for higher dimensions will be available.

The first aspect regards how the critical behavior of the pure gravity theory is modified
by the presence of gauge fields. This is of course relevant to the 
search for a critical point where a renormalized theory of quantum gravity could eventually be
defined, since the coupling to other fields, either matter or gauge, modifies
the renormalization group flow. As we will discuss, the two-dimensional case is somewhat
trivial in this respect, since the coupling to gauge fields just renormalizes the cosmological 
constant and shifts its critical value, leaving critical indices unchanged. 

As a second aspect, we consider the definition and behavior of gauge observables in the fluctuating
space-time geometry. Among various possible gauge invariant quantities (see Ref.~\cite{cdtgauge_pisa1} 
for a more detailed discussion), we focus our attention here on observables related to the
topology of gauge fields, in particular on the so-called winding number (or topological charge): 
this quantity classifies the possible mappings of the gauge group onto the space-time manifold,
therefore it could be particularly interesting in a situation in which the latter 
is not fixed but instead fluctuating.

Before starting the discussion of results, let us briefly recap 
the main features of our simulations. The parameters 
entering the discretized theory are the bare
cosmological constant $\lambda$, see Eq.~\eqref{eq:mincoup-CDT2Daction}, 
and the inverse gauge coupling $\beta \propto 1/g^2$, see
Eq.~\eqref{eq:un-YMact_cdt}, which differs from standard definitions 
used in lattice gauge theories by a proportionality 
factor, related to the bone coordination numbers $n_b$  
and to the fact that the reference flat discretization is hexagonal, 
rather than square, lattice.
A further parameter is the 
number of temporal slices $N_t$, which is fixed, while the spatial 
extension of each slice is dynamical.
The overall imposed topology is that of a torus, 
with periodic boundary conditions in the temporal and spatial 
directions: because of the Gauss-Bonnet theorem in 
two-dimensions, which implies a globally flat geometry, i.e.~the condition
$\langle n_b \rangle = 6$ holds for all sampled triangulations.

\subsubsection{Critical behaviour}

The pure gravity theory undergoes a continuous transition for a critical cosmological parameter
$\lambda_c = \log(2)$. As the critical point is approached from above, 
one observes a divergence both of the average total volume $\langle V \rangle$ of the triangulation and of the 
length $\xi_{\text{Vprof.}}$ describing the large distance behaviour of the correlation
between the spatial extensions of pairs of different time slices, i.e.~the two-point function
of the so-called volume profile. The critical behaviour is described in terms 
of two critical indices $\nu$ and $\mu$:
\begin{equation}\label{eq:un-scalingfunc}
    \langle V \rangle \propto (\lambda - \lambda_c)^{-\mu} \ , \ \ \ \ 
    \xi_{\text{Vprof.}} \propto (\lambda - \lambda_c)^{-\nu} \; .
\end{equation}
Such scaling laws are expected to hold sufficiently close to $\lambda_c$, i.e.~in the so-called
{scaling window}; results of Ref.~\cite{cdtgauge_pisa1} show that the scaling window
for $\xi_{\text{Vprof.}}$ is somewhat larger than that for $\langle V \rangle$.

The addition of gauge fields is not expected to lead to a significant 
modification of the critical behaviour. Indeed,
 it is known that two-dimensional gauge fields can be easily integrated away~\cite{cdtgauge_anal,Cao:2013na,bonati_flatsuscu1}.
  Nevertheless, on a curved geometry one is left
 with a non-trivial contribution, depending on the coordination
 numbers $n_b$, which could modify the behavior of gravity observables.

\begin{figure}[!t]
\centering
\includegraphics[width=0.49\textwidth]{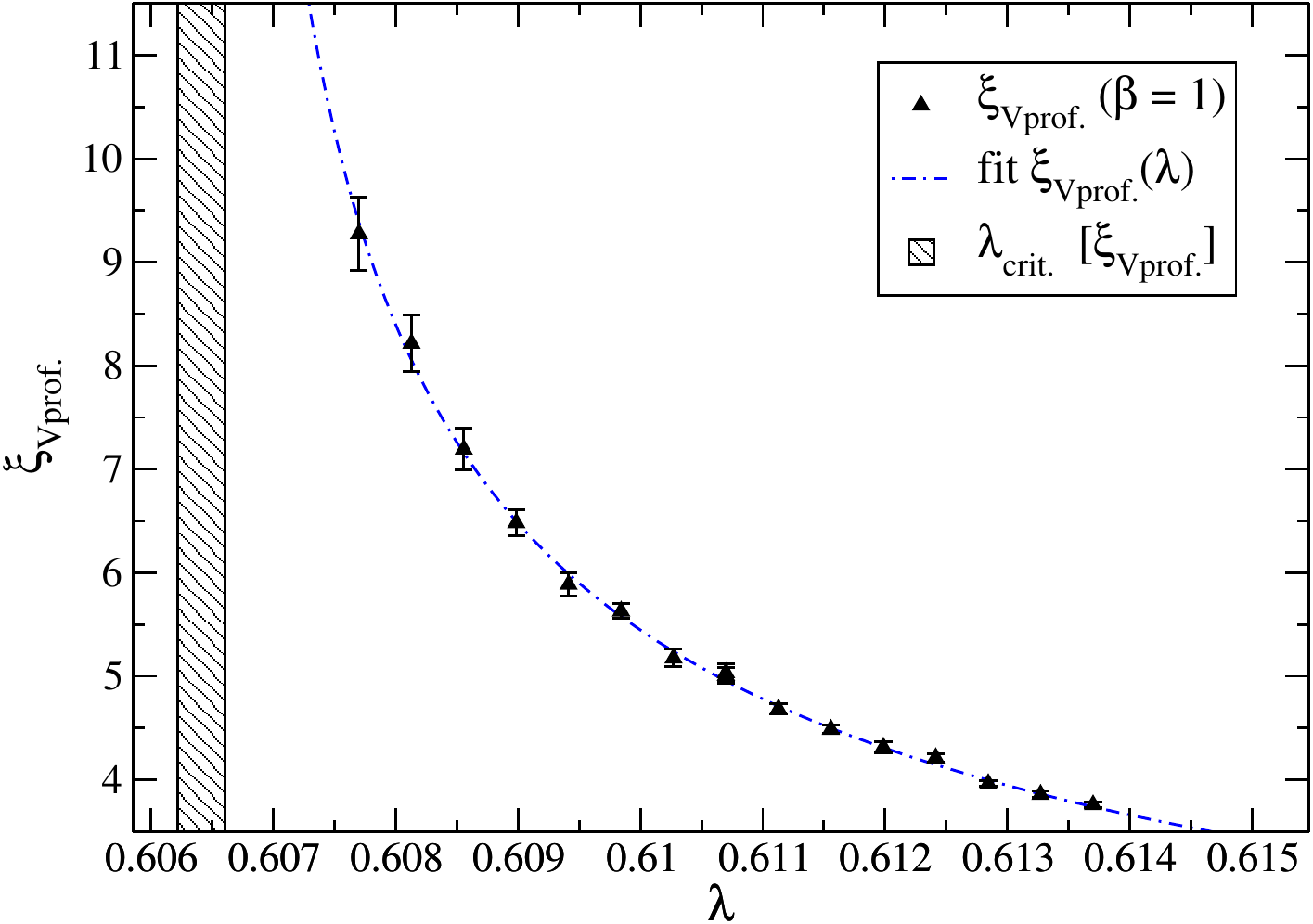}
\includegraphics[width=0.49\textwidth]{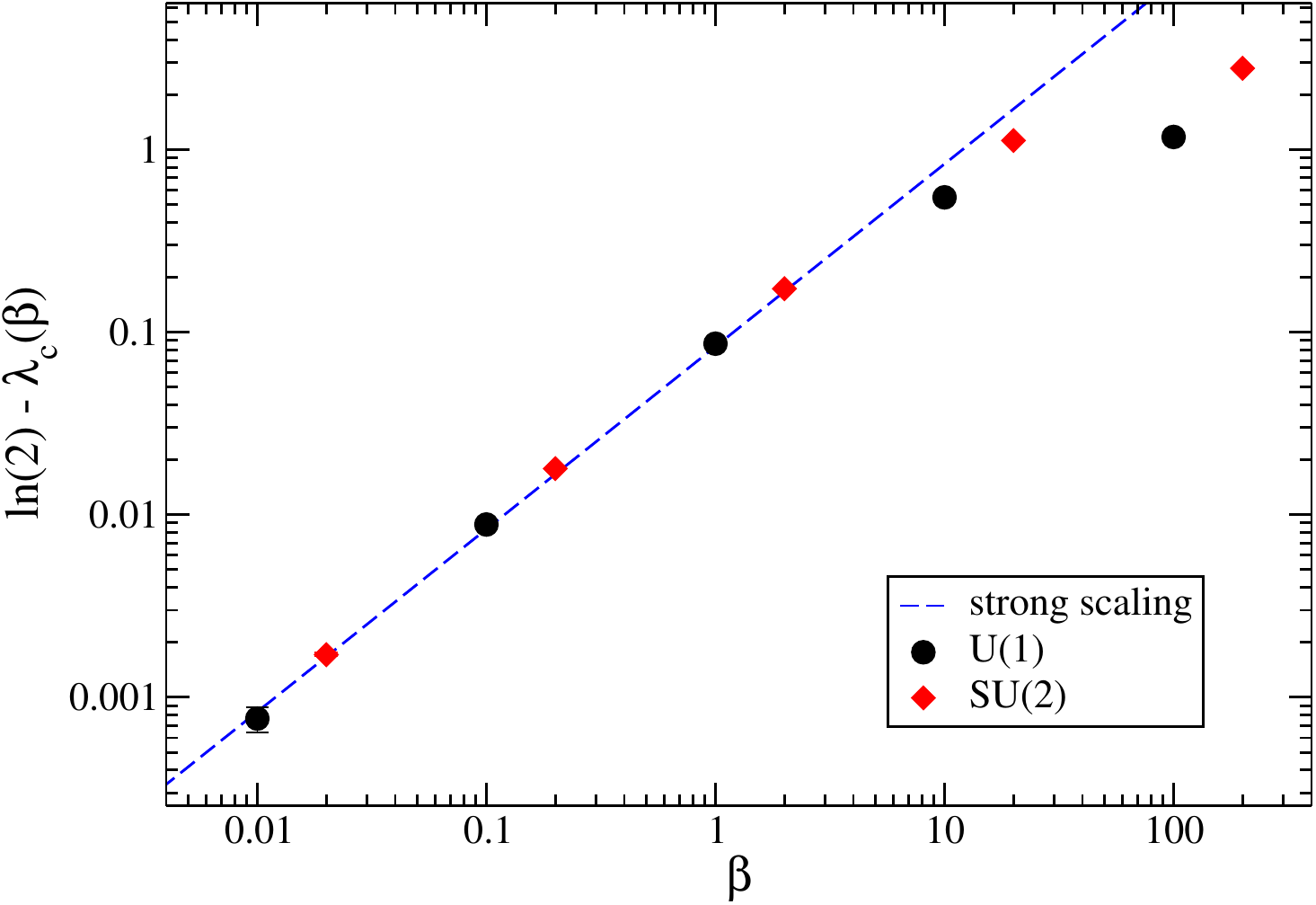}
\caption{Left: determination of 
$\xi_{\text{Vprof.}}$ as a function of $\lambda$ for the $U(1)$
gauge group at $\beta = 1$, together with the results of a fit to a power law behavior
as in Eq.~\eqref{eq:un-scalingfunc}. Right: Critical values of $\lambda$  for $U(1)$ and $SU(2)$ 
as a function of $\beta$; the dashed line represents the strong coupling expansion prediction discussed in the text.}
\label{fig:u1critfit_beta1_onlyxi}
\end{figure}

In Fig.~\ref{fig:u1critfit_beta1_onlyxi} we report the behaviour of the correlation length as a function
of $\lambda$ for $\beta = 1$ and the $U(1)$ gauge group: results can still be nicely fitted according the 
scaling ansatz of Eq.~\eqref{eq:un-scalingfunc}, however with a different $\lambda_c$ compared to the pure
gravity case. The critical values of $\lambda$  obtained for $U(1)$ and $SU(2)$ are reported in the same figure
as a function of $\beta$, together with an analytical prediction based on a strong coupling expansion of the 
gauge theory, i.e.~a series expansion in $\beta$~\cite{cdtgauge_pisa1}. 

\begin{figure}[!t]
\centering
\includegraphics[width=0.9\textwidth]{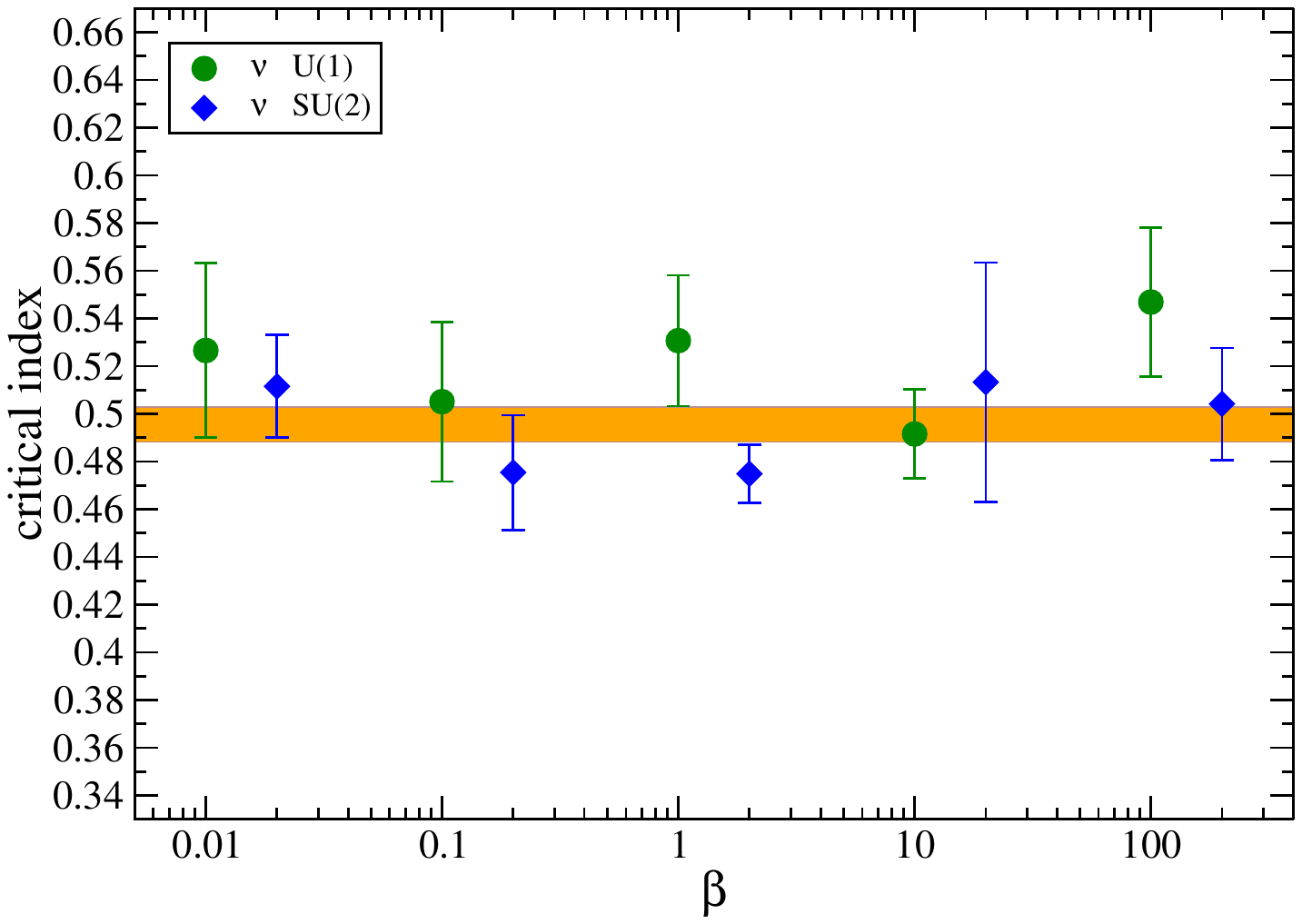}
\caption{Critical index $\nu$, 
as a function of the bare gauge coupling
$\beta$, for both explored gauge groups.
The horizontal bands is the result of a best fit 
to constant value.}
\label{fig:fit_critidx}
\end{figure}

Despite the change in the critical coupling, no appreciable change is observed,
within errors, for the critical indices,
as can be appreciated from Figure~\ref{fig:fit_critidx}, where the values of $\nu$
obtained for different $\beta$ and gauge groups are reported.
To confirm the apparent stability of $\nu$, we have performed a global
fit of all values to a constant, obtaining 
$\nu = 0.496(7)$ ($\chi^2 / \textrm{dof} = 9.1/9$).
We notice that this value is compatible with a mean field 
critical index 1/2.

These observations suggest that, in the two dimensional case, 
the presence of gauge fields modifies gravity just by an additive renormalization
of the cosmological constant. The fact that local fluctuations of $n_b$ are not 
relevant could be explained by the seemingly mean field behaviour. 
Of course, the situation could be quite different in higher dimensions.

As a final comment, it is clear that approaching $\lambda_c$ at fixed 
inverse gauge coupling $\beta$ is not enough to define a proper continuum limit 
for the whole theory. One should tune also $\beta$ so that gauge-related
correlation lengths diverge at the same time: in two space-time dimensions
this is usually achieved as $\beta \to \infty$, with an asymptotic
scaling of gauge correlation lengths proportional to $\sqrt{\beta}$.

\subsubsection{Gauge topology}

In two space-time dimensions, a topological classification of gauge configurations 
applies only to the case of $U(1)$ gauge group. In this case, the topological charge $Q$,
or winding number, amounts to the total flux of the gauge field strength across
the space-time manifold: gauge configurations with a non-integer topological 
charge are completely suppressed from the path-integral if the manifold is compact 
(like in our case, since we are on a torus), or if vanishing conditions at 
infinity are imposed on the field strength, so that relevant contributions to the 
path integral can be classified according to integer values of $Q$. 

The very concept of homotopy classes is lost on a discrete space-time, but is recovered
as the continuum limit is approached. A side effect of that, however, is the fact that,
as the inverse gauge coupling $\beta$ grows,
standard updating algorithms become extremely inefficient in moving from 
one topological sector to the other, so that ergodicity is lost: this problem
is usually known as topological freezing and affects standard simulations
of lattice gauge theories~\cite{Alles:1996vn,top_freeze,Luscher:2011kk,Bonati:2017woi}.

Within our formulation  it is relatively easy to define an integer valued topological charge even on the discretized manifold
\begin{equation}\label{eq:un-topcharge_U1}
	Q \equiv \frac{1}{2 \pi}\sum_{b\in \mcTau^{(d-2)}} \arg\lbrack Tr (\Pi_b)\rbrack,
\end{equation}
where $\Pi_b$ is the plaquette around vertex $b$, 
and the argument function $\arg z$ returns values in $(-\pi,\pi\rbrack$.

\begin{figure}[!t]
\centering
\includegraphics[width=0.9\textwidth]{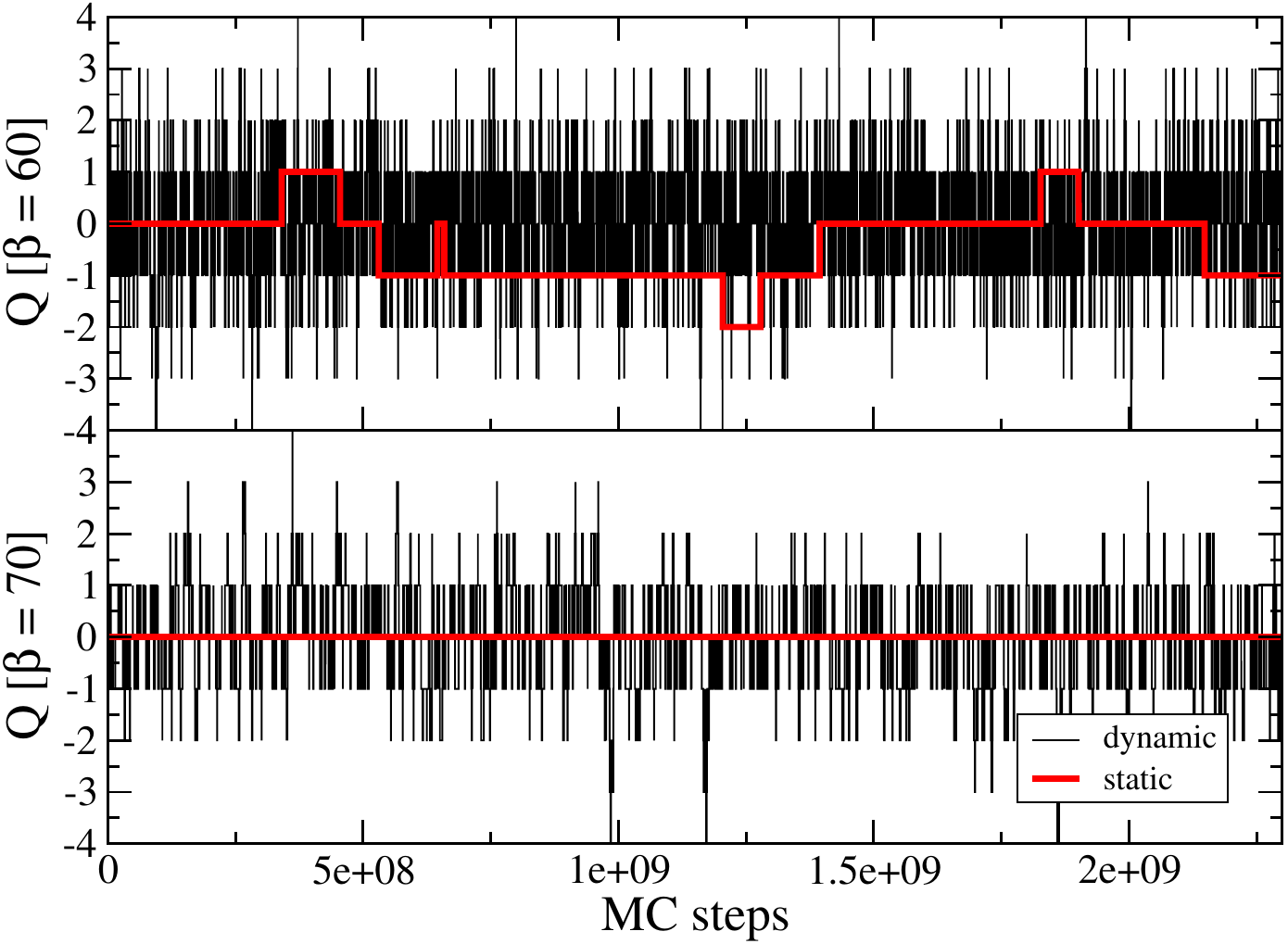}
\caption{Comparison of the Monte-Carlo histories of $Q$ 
    for the dynamic and static (flat) simulations with a comparable total volume
    $\langle V \rangle = 800$ and for two different inverse gauge couplings, $\beta = 60$ and $\beta = 70$.}
\label{fig:Qhistories_compare}
\end{figure}

In the following we discuss briefly results obtained for $Q$ and for the so-called
topological susceptibility $\chi \equiv \langle Q^2 \rangle / V$,
comparing results from numerical simulations on a triangulation with fixed 
geometry, corresponding in particular to a flat torus, with those from 
dynamical simulations in which the geometry is updated as well.
To start with, in
Figure~\ref{fig:Qhistories_compare} we compare 
the Monte Carlo histories of $Q$
obtained in the two cases for $\beta=60$ and $\beta = 70$.
In order to make 
the comparison meaningful, we tuned $\lambda$ so has to have an average volume $\langle V \rangle$ equal to 
the volume used for the fixed geometry case.

Topological freezing emerges quite clearly in static simulations, indeed
$Q$ is fully frozen at $\beta = 70$; the interesting result is that, on the contrary,
no freezing at all is observed in the dynamical case. 
We have checked that this phenomenon is induced  by the typical roughness of the triangulation,
rather than by its dynamical change during the Monte-Carlo evolution (see Ref.~\cite{cdtgauge_pisa1} for details).
A possible explanation of this substantial improvement in the decorrelation of $Q$ could be searched in the
existence of regions with a large negative curvature (i.e., of vertices with a large coordination number $n_b$),
where large fluctuations in the local flux of the field strength across the manifold are possible, with 
a limited expense in terms of the pure gauge action. It is reasonable to guess that this feature could extend to the 
higher dimensional case as well.

\begin{figure}[!t]
\centering
\includegraphics[width=0.9\textwidth]{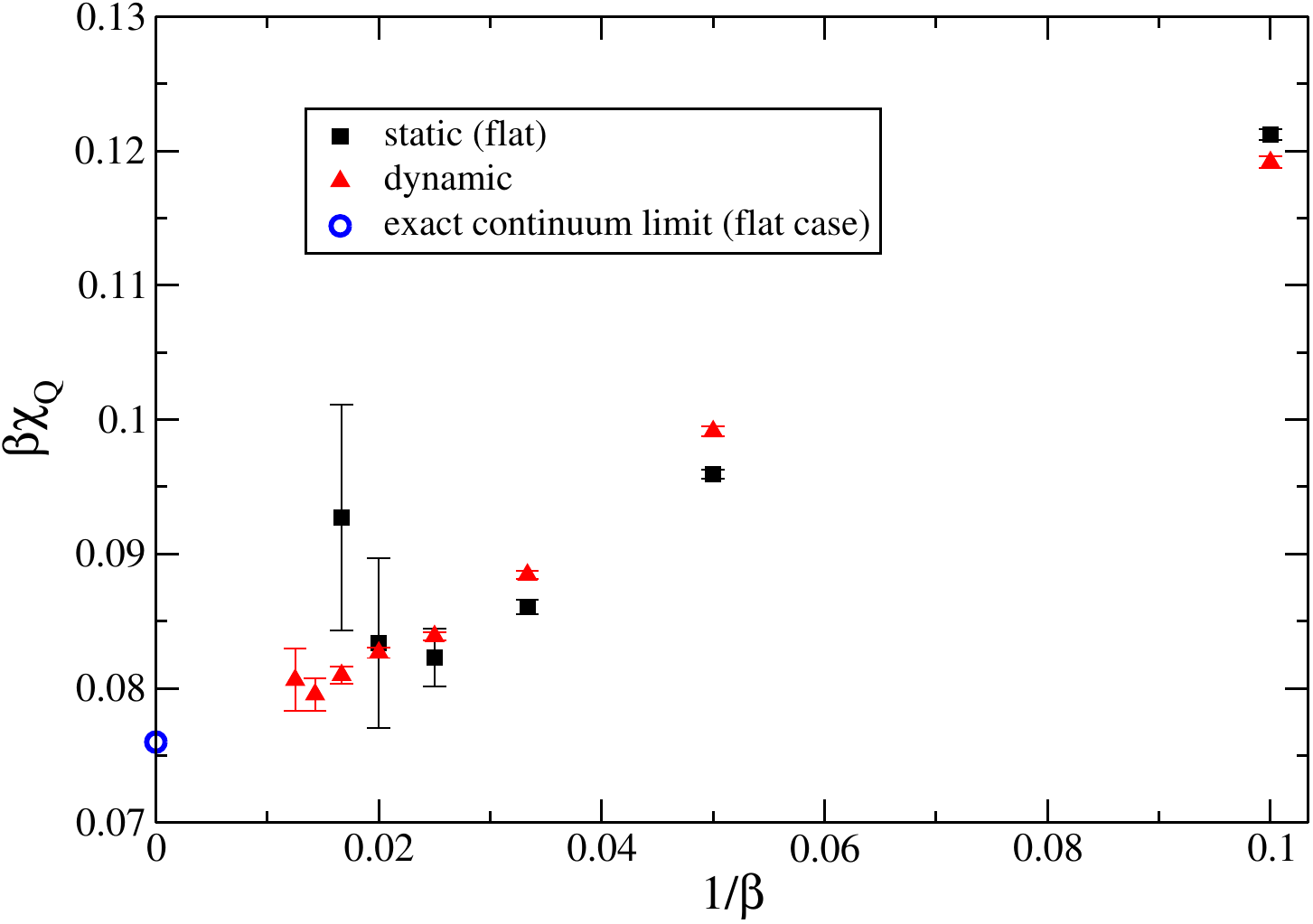}
\caption{Comparison of the topological susceptibilities $\langle Q^2 / V \rangle$ obtained for the static (flat) and dynamic 
simulations at average volume $V=800$.}
\label{fig:topsusc_stat_dyn_20x20}
\end{figure}

To conclude, in
Figure~\ref{fig:topsusc_stat_dyn_20x20} we show results obtained for the topological susceptibility 
(times $\beta$), comparing again
the flat static and the dynamical case. We stress that, in the dynamical case,
an ambiguity emerges in the definition of $\chi$, since the total volume itself fluctuates;
our prescription is to define
$\chi_Q \equiv \langle \frac{Q^2}{V}\rangle$,
i.e., volume fluctuations are taken into account and included in one single average.
Slight differences are observed at finite $\beta$,  
however, the improved decorrelation of $Q$ in the dynamical case
allows to obtain accurate results even at large values of $\beta$.
Assuming $1/\beta$ corrections and considering only data
with $\beta > 30$, we have tried to extract a continuum
extrapolation (however, at fixed gravity coupling) obtaining 
$\beta \chi_Q = 0.0758(14)$ 
with a reduced chi-squared $\chi^2/\textrm{dof} = 1.6/3$: 
such result is in good agreement with the 
prediction for the flat continuum theory~\cite{bonati_flatsuscu1}, 
 $\beta \chi_Q = 3 / (4 \pi^2)\simeq 0.075991$, where we have
taken into account an additional factor 6 in our definition 
of $\beta$.

\section{Summary}
\label{sec:conclusions}

This Chapter has been devoted to a few recent developments regarding the CDT approach to Quantum Gravity. 
As a first topic, in Section~\ref{sec:spectral_methods} we have discussed the need for new physical quantities characterizing the CDT phase diagram,
focusing on spectral observables. 
We showed how information about the effective dimension at large scales 
and the characteristic scales of ensembles of geometries
can be extracted from the spectrum, and discussed another 
possible discretization. However, there is still much to be investigated
in the form of analysis of the properties of eigenvectors, 
such as their localization properties 
or a coarse-graining procedure for local observables by spectral projection
in the eigenspace corresponding to the lowest part of the spectrum. 

As a second topic, we have discussed how the CDT approach could be enlarged 
to include a minimal coupling to Abelian and non-Abelian gauge fields. While the general formulation 
is clear, present algorithms limit the exploration to the two-dimensional case. As a preliminary step,
which is possible already with available algorithms, one could perform numerical simulations of 
gauge theories on fixed triangulations sampled via pure gravity CDT simulations: that would not give any
information on the feedback of gauge fields on gravity, however could reveal interesting aspects about
the influence of the underlying space-time geometry on gauge field dynamics.

%
%

\begin{thebibliography}{99.}%

\bibitem{Ambjorn_chapt1}
J. Ambj{\o}rn, 
``(Causal) Dynamical Triangulations: a Regularization of Quantum Gravity'',
Chapter 1 of the Section ``Causal Dynamical Triangulations'' of the ``Handbook of Quantum Gravity'' (Eds. C. Bambi, L. Modesto and I.L. Shapiro, Springer Singapore, expected in 2023).

\bibitem{Gizbert-Studnicki_chapt10}
J. Gizbert-Studnicki,
``Semiclassical and Continuum Limits of Four-Dimensional CDT'',
Chapter 10 of the Section ``Causal Dynamical Triangulations'' of the ``Handbook of Quantum Gravity'' (Eds. C. Bambi, L. Modesto and I.L. Shapiro, Springer Singapore, expected in 2023).


\bibitem{reuter_cad}
 M.~Reuter, F.~Wolter, M.~Shenton and M.~Niethammer,
 ``Laplace--Beltrami Eigenvalues and Topological Features of Eigenfunctions for Statistical Shape Analysis,''
 Computer-Aided Design {\bf 41} no.10, 739 (2009)

 \bibitem{reuter_dna}
 M.~Reuter, F.~Wolter, M.~Shenton and M.~Niethammer,
 Computer-Aided Design {\bf 38} no.4, 342 (2006)
 

\bibitem{lapl_embedding}
M.~Belkin and P.~Niyogi. 
``Laplacian eigenmaps and spectral techniques for embedding and clustering,''
In Proceedings of NIPS’01 (2001) 585–591.

\bibitem{heatk}
D.V.~Vassilevich, 
``Heat kernel expansion: user's manual'',
Phys. Rept. {\bf 388}, 5 (2003).


\bibitem{Anderson:1958vr}
P.~W.~Anderson,
``Absence of Diffusion in Certain Random Lattices''
Phys. Rev. \textbf{109} (1958), 1492-1505

\bibitem{qcd_anderson_multifractal}
  L.~Ujfalusi, M.~Giordano, F.~Pittler, T.~G.~Kovacs and I.~Varga,
  ``Anderson transition and multifractals in the spectrum of the Dirac operator of Quantum Chromodynamics at high temperature,''
  Phys.\ Rev.\ D {\bf 92} (2015) no.9,  094513
  [arXiv:1507.02162 [cond-mat.dis-nn]].

\bibitem{fractals_havlinbook}
    D.~ben-Avraham and S.~Havlin,
    ``Diffusion and Reactions in Fractals and Disordered Systems'',
    Cambridge University Press (2005)
    
\bibitem{edt_spectral_dim}
J.~Ambj{\o}rn, D.~Boulatov, J.L.~Nielsen, J.~Rolf and Y.~Watabiki, 
    ``The spectral dimension of 2-D quantum gravity'',
    JHEP {\bf 02} (1998) 010 
    arXiv:hep-th/9801099.

\bibitem{cdt_spectral_dim}
    J.~Ambj{\o}rn, J.~Jurkiewicz and R.~Loll,
    ``The Spectral Dimension of the Universe is Scale Dependent'',
    Phys.\ Rev.\ Lett. {\bf 95} (2005) 171301
    arXiv:hep-th/0505113
    
\bibitem{diffproc}
	J.~Ambjorn, J.~Jurkiewicz and R.~Loll,
	``Reconstructing the universe''
	Phys.\ Rev.\ D {\bf 72} (2005) 064014

\bibitem{heatrace_coeffs}
 H.~P.~McKean, I.~Singer,
``Curvature and the eigenvalues of the Laplacian'',
 J. Differential Geometry {\bf 1} no.1, 43 (1967)

 \bibitem{weylslaw_1}
H.~Weyl,
Nachr. K\"onigl. Ges. Wiss. G\"ottingen, 110--117 (1911). 
 
 \bibitem{weylslaw}
V.~Ivrii,
``100 years of Weyl's law'',
Bulletin of Mathematical Sciences {\bf 6} (2016) no.3, 379--452
[arXiv:1608.03963v2 [math.SP]].

\bibitem{LBseminal}
G.~Clemente and M.~D'Elia,
``Spectrum of the Laplace-Beltrami operator and the phase structure of causal dynamical triangulations'',
\emph{Phys. Rev. D} \textbf{97} (2018) 124022
[arXiv:1804.02294 [hep-th]].

\bibitem{cheeger}
J.~Cheeger,
``A lower bound for the smallest eigenvalue of the Laplacian,''
Problems in analysis (Sympos. in honor of Salomon Bochner, Princeton Univ., Princeton, N.J., 1969), pp. 195–199. Princeton Univ. Press, Princeton, N.J., 1970.

\bibitem{cdt_secondord}
J.~Ambjorn, S.~Jordan, J.~Jurkiewicz and R.~Loll,
``A Second-order phase transition in CDT'',
\emph{Phys. Rev. Lett.} {\bf 107} (2011) 211303
[arXiv:1108.3932 [hep-th]].

\bibitem{cdt_secondfirst}
J.~Ambjorn, S.~Jordan, J.~Jurkiewicz and R.~Loll,
``Second- and First-Order Phase Transitions in CDT,''
\emph{Phys. Rev. D} {\bf 85} (2012) 124044
[arXiv:1205.1229 [hep-th]].

\bibitem{new_phase_chars}
J.~Ambj\o{}rn, J.~Gizbert-Studnicki, A.~G\"orlich, J.~Jurkiewicz, N.~Klitgaard and R.~Loll,
``Characteristics of the new phase in CDT,''
\emph{Eur. Phys. J. C} {\bf 77} (2017) 152
[arXiv:1610.05245 [hep-th]].

\bibitem{cdt_newhightrans}
J.~Ambjorn, D.~Coumbe, J.~Gizbert-Studnicki, A.~Gorlich and J.~Jurkiewicz,
``New higher-order transition in causal dynamical triangulations,''
\emph{Phys. Rev. D} \textbf{95} (2017) 124029
[arXiv:1704.04373 [hep-lat]].

 \bibitem{cdt_desitter}
 J.~Ambjorn, J.~Jurkiewicz and R.~Loll,
 Phys.\ Rev.\ Lett.\  {\bf 93} (2004) 131301
 [hep-th/0404156].

\bibitem{Reitz:2022dbj}
M.~Reitz, D.~N\'emeth, D.~Rajbhandari, A.~G\"orlich and J.~Gizbert-Studnicki,
``Generalised spectral dimensions in non-perturbative quantum gravity,''
[arXiv:2207.05117 [gr-qc]].

\bibitem{LBFEMseminal}
F.~Caceffo and G.~Clemente,
``Spectral analysis of causal dynamical triangulations via finite element method,''
Phys. Rev. D \textbf{107} (2023) no.7, 074501
[arXiv:2010.07179 [hep-lat]].

\bibitem{fem_allairebook}
    G.~Allaire and A.~Craig, 
    ``Numerical Analysis and Optimization,'' 
    Oxford University Press (2007)

\bibitem{fem_hughesbook}
    T.J.R.~Hughes,
   ``The Finite Element Method: Linear Static and Dynamic Finite Element Analysis'',
   Dover Publications (2000)

\bibitem{fem_strangbook}
    G.~Strang and G.~Fix,
    ``An Analysis of the Finite Element Method'',
    Wellesley-Cambridge Press (2008)

\bibitem{fem_taylorbook}
    O.C.~Zienkiewicz, R.L.~Taylor and J.Z.~Zhu,
    ``The Finite Element Method: its Basis and Fundamentals'',
    Butterworth-Heinemann (2013)

\bibitem{fem_babuska}
    B.~Szab\'o and I.~Babuska,
   ``Finite Element Analysis'',
   Wiley-Interscience (1991)

\bibitem{LBrunning}
G.~Clemente, M.~D'Elia and A.~Ferraro,
``Running scales in causal dynamical triangulations,''
Phys. Rev. D \textbf{99} (2019) no.11, 114506
[arXiv:1903.00430 [hep-th]].

\bibitem{lbpos19}
    G.~Clemente, M.~D'Elia and A.~Ferraro,
    ``Spectral Methods in Causal Dynamical Triangulations''
    PoS {\bf Lattice2019} 116
    arXiv:1912.11311 [hep-lat]

\bibitem{Wilson:1974sk}
K.~G.~Wilson,
``Confinement of Quarks,''
Phys. Rev. D \textbf{10} (1974), 2445-2459

\bibitem{regge} 
T.~Regge,
``General Relativity Without Coordinates,''
\emph{Nuovo Cim.} {\bf 19} (1961) 558.

\bibitem{cdt_review12} 
J.~Ambjorn, A.~Goerlich, J.~Jurkiewicz and R.~Loll,
``Nonperturbative Quantum Gravity,''
\emph{Phys.\ Rept.}  {\bf 519} (2012) 127
[arXiv:1203.3591 [hep-th]].

\bibitem{metro}
N.~Metropolis, A.~W.~Rosenbluth, M.~N.~Rosenbluth, A.~H.~Teller and E.~Teller,
``Equation of state calculations by fast computing machines,''
\emph{J. Chem. Phys.} \textbf{21} (1953) 1087

\bibitem{hastings}
W.~K.~Hastings,
``Monte Carlo Sampling Methods Using Markov Chains and Their Applications,''
\emph{Biometrika} \textbf{57} (1970) 97

\bibitem{alexander}
J.W. Alexander,
``The combinatorial theory of complexes,''
\emph{Ann. Mat.} {\bf 31} (1931) 292.

\bibitem{cdt_2dmoves}
J.~Ambj\o{}rn, J.~Jurkiewicz and R.~Loll,
``Lorentzian and Euclidean Quantum Gravity \textemdash{} Analytical and Numerical Results,''
\emph{NATO Sci. Ser. C} \textbf{556} (2000) 381
[arXiv:hep-th/0001124 [hep-th]].

\bibitem{cdtgauge_pisa1}
A.~Candido, G.~Clemente, M.~D'Elia and F.~Rottoli,
``Compact gauge fields on Causal Dynamical Triangulations: a 2D case study,''
JHEP \textbf{04} (2021), 184
[arXiv:2010.15714 [hep-lat]].

\bibitem{hb_creutz}
M.~Creutz,
``Monte Carlo Study of Quantized SU(2) Gauge Theory,''
\emph{Phys. Rev. D} \textbf{21} (1980) 2308

\bibitem{hb_kennedy-pendleton}
A.~D.~Kennedy and B.~J.~Pendleton,
``Improved Heat Bath Method for Monte Carlo Calculations in Lattice Gauge Theories,''
\emph{Phys. Lett. B} \textbf{156} (1985) 393

\bibitem{Cabibbo:1982zn}
N.~Cabibbo and E.~Marinari,
``A New Method for Updating SU(N) Matrices in Computer Simulations of Gauge Theories,''
Phys. Lett. B \textbf{119} (1982), 387-390
doi:10.1016/0370-2693(82)90696-7

\bibitem{cdtgauge_anal}
J.~Ambjorn and A.~Ipsen,
``Two-dimensional causal dynamical triangulations with gauge fields,''
Phys. Rev. D \textbf{88} (2013) no.6, 067502
[arXiv:1305.3148 [hep-th]].

\bibitem{Cao:2013na}
C.~Cao, M.~van Caspel and A.~R.~Zhitnitsky,
``Topological Casimir effect in Maxwell Electrodynamics on a Compact Manifold,''
\emph{Phys. Rev. D} \textbf{87} (2013) 105012
[arXiv:1301.1706 [hep-th]].

\bibitem{bonati_flatsuscu1}
C.~Bonati and P.~Rossi,
``Topological susceptibility of two-dimensional $U(N)$ gauge theories,''
\emph{Phys. Rev. D} \textbf{99} (2019) 054503
[arXiv:1901.09830 [hep-lat]].

\bibitem{Alles:1996vn}
B.~Alles, G.~Boyd, M.~D'Elia, A.~Di Giacomo and E.~Vicari,
``Hybrid Monte Carlo and topological modes of full QCD,''
\emph{Phys. Lett. B} \textbf{389} (1996) 107
[arXiv:hep-lat/9607049 [hep-lat]].

\bibitem{top_freeze}
L.~Del Debbio, G.~M.~Manca and E.~Vicari,
``Critical slowing down of topological modes,''
\emph{Phys. Lett. B} \textbf{594} (2004) 315
[arXiv:hep-lat/0403001 [hep-lat]].

\bibitem{Luscher:2011kk}
M.~Luscher and S.~Schaefer,
``Lattice QCD without topology barriers,''
\emph{JHEP} \textbf{07} (2011) 036
[arXiv:1105.4749 [hep-lat]].


\bibitem{Bonati:2017woi}
C.~Bonati and M.~D'Elia,
``Topological critical slowing down: variations on a toy model,''
\emph{Phys. Rev. E} \textbf{98} (2018) 013308
[arXiv:1709.10034 [hep-lat]].



%
%
%
%
%
%
%
%
%
%
%
%
%
%
%
%
%
%
%
%
%
%
%
%
%
%
%
%
%
%
%
%
%
%
%
%
%
%
%
%
%
%
%
%
%
%
%
%
%
%

%
%
%
%
%
%
%
%
%
%
%
%
%
%
%
%
%
%
%
%
%
%
%
%
%
%
%
%
%
%
%
%
%
%
%
%
%
%
%
%
%
%
%
%
%
%
%
%
%
%
%
%
%


 
\end{thebibliography}
%

\end{document}